\documentclass[journal]{IEEEtran}

\usepackage{cite}
\usepackage{amsbsy,amsmath,amssymb,epsfig,bbm,mathrsfs,multirow,amsthm,amsfonts}
\usepackage{algorithm,algorithmic}
\usepackage{graphicx}
\usepackage{textcomp}
\usepackage{multicol}
\usepackage{float}
\usepackage{url}
\usepackage{rotating}
\usepackage{enumitem}
\newlist{tabitemize}{itemize}{1}
\setlist[tabitemize]{leftmargin=0.25cm,
                     label      = $\bullet$ ,
                     after      = \vspace{-\baselineskip}}
\usepackage[table]{xcolor}

\PassOptionsToPackage{hyphens}{url}
\graphicspath{{./figures/}}

\def\BibTeX{{\rm B\kern-.05em{\sc i\kern-.025em b}\kern-.08em
    T\kern-.1667em\lower.7ex\hbox{E}\kern-.125emX}}

\newtheorem{definition}{Definition}

\usepackage[T1]{fontenc}
\usepackage{makecell}
\usepackage{amstext} 
\usepackage{subcaption} 
\definecolor{mygray}{gray}{0.6}
\definecolor{myblue}{rgb}{0.8,0.85,1}

\usepackage{array, tabularx, boldline}
\newcolumntype{L}[1]{>{\raggedright\let\newline\\\arraybackslash\hspace{0pt}}m{#1}}
\newcolumntype{C}[1]{>{\centering\let\newline\\\arraybackslash\hspace{0pt}}m{#1}}
\newcolumntype{R}[1]{>{\raggedleft\let\newline\\\arraybackslash\hspace{0pt}}m{#1}}

\usepackage{cellspace}
\begin{document}

\title{Federated Unlearning in Edge Networks: A Survey of Fundamentals, Challenges, Practical Applications and Future Directions}

\author{
Jer Shyuan Ng,
Wathsara Daluwatta,
Shehan Edirimannage, 
Charitha Elvitigala,\\
Asitha Kottahachchi Kankanamge Don,
Ibrahim Khalil,
Heng Zhang, 
Dusit Niyato
%~\thanks{Jer Shyuan Ng, Wathsara Daluwatta, Shehan Edirimannage, Charitha Elvitigala, Asitha Kottahachchi Kankanamge Don and Ibrahim Khalil are with the School of Computing Technologies, RMIT University, Melbourne, VIC 3000, Australia (e-mail: jer.shyuan.ng@rmit.edu.au, wathsara.daluwatta@student.rmit.edu.au; shehan.edirimannage@student.rmit.edu.au; charitha.elvitigala@student.rmit.edu.au; asitha.kottahachchi.kankanamge.don@student.rmit.edu.au; ibrahim.khalil@rmit.edu.au).}
%~\thanks{Heng Zhang is with the School of Journalism & Communication, Chonqing University, China (email:hengzhang@cqu.edu.cn)}
%~\thanks{Dusit Niyato is with the College of Computing and Data Science, Nanyang Technological University, Singapore 639798 (e-mail: dniyato@ntu.edu.sg).}
}

\maketitle

\begin{abstract}
The proliferation of connected devices and privacy-sensitive applications has accelerated the adoption of Federated Learning (FL), a decentralized paradigm that enables collaborative model training without sharing raw data. While FL addresses data locality and privacy concerns, it does not inherently support data deletion requests that are increasingly mandated by regulations such as the Right to be Forgotten (RTBF). In centralized learning, this challenge has been studied under the concept of Machine Unlearning (MU), that focuses on efficiently removing the influence of specific data samples or clients from trained models. Extending this notion to federated settings has given rise to Federated Unlearning (FUL), a new research area concerned with eliminating the contributions of individual clients or data subsets from the global FL model in a distributed and heterogeneous environment. In this survey, we first introduce the fundamentals of FUL. Then, we review the FUL frameworks that are proposed to address the three main implementation challenges, i.e., communication cost, resource allocation as well as security and privacy. Furthermore, we discuss applications of FUL in the modern distributed computer networks. We also highlight the open challenges and future research opportunities. By consolidating existing knowledge and mapping open problems, this survey aims to serve as a foundational reference for researchers and practitioners seeking to advance FL to build trustworthy, regulation-compliant and user-centric federated systems.
\end{abstract}


\section{Introduction}

Traditionally, data is aggregate on a central server where Artificial Intelligence (AI) models are trained in a centralized fashion. However, with the rapid proliferation of data generated by edge devices such as laptops, smartphones, mobile platforms and Internet-of-Things (IoT) systems, this centralized approach has become increasingly impractical. The centralized training method faces three main limitations: (i) high communication overhead, as large volumes of data must be transmitted over the communication networks frequently, (ii) computational inefficiency, since training complex AI models on massive datasets requires substantial centralized resources, (iii) privacy risks, as transferring sensitive user data to a central repository heightens the risk of data breaches and unauthorized access. These challenges have motivated the paradigm shift from centralized machine learning toward a decentralized training paradigm, known as Federated Learning (FL).

In contrast to conventional centralized learning, where raw data is collected and stored in a central repository, FL enables multiple clients to train a shared global FL model collaboratively while retaining their data locally on their devices. In particular, the FL server first distributes the global FL model to the participating clients. Given the global FL model, each client then updates the model using its local dataset, producing new local model parameters. These updated local model parameters are transmitted to the FL server that aggregates them using algorithms such as \emph{FedAvg}~\cite{mcmahan2017communication}. The updated global FL model is subsequently redistributed to clients for the next round of training. This iterative process continues until the model converges to the desired level of accuracy. 

Building on the advantages of FL, i.e., low latency, communication-efficient and privacy-preserving, FL has demonstrated notable success in practical applications. For example, it has been employed to improve the next-word prediction model in Google's Gboard~\cite{hard2019keyboard}. Beyond mobile services, FL has been increasingly adopted in privacy-sensitive domains such as finance~\cite{long2020banking}, healthcare~\cite{chaddad2024healthcare}, transportation~\cite{xu2023transportation} and smart cities~\cite{ragab2025smartcities}. For example, an FL-trained clinical decision support model was developed to predict oxygen requirements in symptomatic COVID-19 patients using multi-modal data inputs such as laboratory data, chest radiographs and vital signs, thereby reducing reliance on operator-dependent and labour-intensive assessments~\cite{dayan2021clinical}. Similarly, in the finance sector, FL has been applied to model custormer behavior, such as predicting loan default risks, while safeguarding sensitive customer data~\cite{shingi2020loan}.

Although FL offers a strong foundation for privacy preservation, the challenges of data deletion and user revocation have gained increasing importance. Modern systems are expected not only to secure data during training but also to comply with evolving user rights and regulatory requirements. For example, the Right to be Forgotten (RTBF), established in frameworks such as the General Data Protection Regulation (GDPR) and the Health Insurance Portability and Accountability Act (HIPAA), grants individuals the right to request the removal of their data from service providers. In the context of machine learning, this has spurred interest in Machine Unlearning (MU). Simply deleting stored user data is not sufficient as its influence persists in models that were trained on it. MU addresses this gap by ensuring that a model behaves as if specific data samples, clients or updates had never contributed to the training process, thus aligning storage-level deletion with model-level forgetting. The most straightforward method is full retraining, where the data or clients to be removed are excluded from the training process. However, retraining is computationally expensive and often infeasible for large scale models, e.g, Large Language Models (LLMs), particularly when unlearning must be performed repeatedly upon new deletion requests. Consequently, MU focuses on developing efficient techniques to remove the influence of unwanted data while maintaining model utility, i.e., model accuracy. 

Extending the concept of unlearning to federated settings has given rise to Federated Unlearning (FUL). However, due to the decentralized, heterogeneous and dynamic characteristics of FL, conventional machine unlearning strategies cannot be directly adopted. First, unlike centralized training, where data resides on a single server, FL involves a diverse set of clients with non-IID and constantly evolving data distributions as well as varying resource capacities. The removal of a client's contribution can disproportionately affect the global model, particularly if the client holds rare or unique data patterns. Second, FL inherently features dynamic participation, where clients may join or leave the federation at any time. This complicates unlearning, as requests from clients that have dropped out are difficult to manage and ensuring that past aggregated model updates from such clients can be effectively "forgotten" in subsequent rounds remain a major hurdle. Third, the FL server has only limited visibility of the training data, since it only receives parameter updates and never directly accesses raw client data. Finally, FUL is vulnerable to adversarial misuse, where malicious clients may issue unlearning requests with the intent to weaken the model or introduce fairness concerns, e.g., by eliminating data from specific classes essential for robust generalization.

Given that FUL relies on repeated communication between clients and FL server, its feasibility and performance are highly sensitive to network dynamics. In particular, bandwidth constraints are a primary limiting factor. Unlearning often requires corrective updates, rollback operations or recovery rounds that introduce additional communication beyond standard FL training. Thus, FUL methods require compact updates and localized participation to reduce network load. Besides, security and reliability of communication further influcence FUL. Attackers may exploit deletion phases by injecting poisoned updates or replaying stale information. Secure aggregation, authentication and consistency checks must therefore remain active during unlearning and recovery, even under partial participation and packet loss. Moreover, the topology and hierarchy of the network also shape FUL behavior. Edge deployments often follow clustered or hierarchical architectures. Unlearning that propagates globally across tiers can incur unnecessary network traffic. Network-aware FUL instead confines deletions to affected clusters or edge regions, allowing unaffected parts of the network to continue normal operations. Therefore, network factors determine \emph{where, when,} and \emph{how much} unlearning can be performed. Effective FUL at the edge requires tight co-design between learning algorithms and networking mechanisms to ensure predictable latency, bounded communication cost and robustness to disconnections. 

The advancement of FUL is essential to ensure that FL systems remain reliable, legally-compliant and responsive to practical requirements. By facilitating the removal of user data contributions when requested, FUL enhances user confidence, promotes ethical and responsible AI practices and broadens the applicability of FL in domains where stringent privacy protections are indispensable.

\subsection{Key Contributions and Comparison}

Although numerous surveys exist on FL and MU, these topics are largely examined in isolation. On one hand, surveys on FL typically overlook the notion of unlearning. On the other hand, surveys on MU seldom address its integration or applicability within federated settings. An overview of the selected surveys in FL and MU is presented in Table~\ref{tab:flmusurvey}.

\begin{table*}[!ht]
\caption{\small An overview of selected surveys in FL and MU.} 
\label{tab:flmusurvey}
\centering
%\begin{tabular}{|l |l| l| l|}
\begin{tabular}{|>{\centering\arraybackslash}m{1cm}|>{\centering\arraybackslash}m{0.6cm} | m{13cm}|}
%{0.75\textwidth}
\hline

\rowcolor{mygray}
\textbf{Subject}  & \textbf{Ref.}  & \multicolumn{1}{c|}{\textbf{Key Contributions}} \\ 
\hline

\multirow{3}{*}{FL} & \cite{wen2023flsurvey}  & Survey on the FL frameworks that address the implementation challenges of FL \\
\cline{2-3}
&   \cite{nguyen201fliot} & Survey on the integration of FL and IoT as well as the emerging applications of FL in the IoT networks\\
\cline{2-3}
& \cite{yuan2024decentralized} & Survey on decentralized FL from different perspectives and its extended variants\\
\hline

\multirow{3}{*}{MU} & \cite{shaik2025mu}  & Detailed taxonomy on various MU techniques and the evaluation methods to assess these MU techniques  \\
\cline{2-3}
&   \cite{liu2025threats} & Survey on the threats, attacks and defenses in MU\\
\cline{2-3}
& \cite{li2025metrics} & Comprehensive review on the existing MU algorithms as well as the verification and evaluation metrics in MU \\
\cline{2-3}
& \cite{qu2024insights} & In-depth discussion on the difference between exact and approximate MU\\
\hline

\end{tabular}
\end{table*}

Existing surveys on FL primarily concentrate on the design of FL frameworks that facilitate the FL training process. In~\cite{wen2023flsurvey}, the authors classify FL frameworks according to key implementation challenges, including communication overhead, system heterogeneity as well as privacy and security concerns. The authors in~\cite{nguyen201fliot} provide a comprehensive overview of the emerging FL applications in IoT networks, emphasizing its potential as an enabling technology for IoT services such as data offloading and caching, data sharing, mobile crowdsensing and attack detection. Beyond the distributed nature of FL, the authors in~\cite{yuan2024decentralized} focus on decentralized FL, where FL model aggregation occurs without reliance on a central server. Their survey discusses decentralized frameworks from multiple dimensions, including iteration order, communication protocols, network topologies, paradigm proposals and temporal variability. While these studies provide valuable insights into the deployment and implementation of FL across edge networks, they largely overlook an equally critical dimension, i.e., the ability to remove the contributions of specific data samples, clients or classes.

\begin{table}[t]
\scriptsize
  \caption{\small Comparison between Exact and Approximate Unlearning~\cite{li2025metrics}.}
  \label{tab:exactapproximate}
  \centering
  %\begin{tabularx}{8.7cm}{|Sl|X|X|}
  \begin{tabular}{|>{\centering\arraybackslash}m{1.1cm}|m{3cm}|m{3cm}|}
  \hline
  Types & \multicolumn{1}{c|}{\textbf{Exact Unlearning}} & \multicolumn{1}{c|}{\textbf{Approximate Unlearning}} \\   \hline

Aims & Aligns the unlearned model's distribution with that of a retrained model & Obtain indistinguishable parameters between the unlearned and retrained model \\ \hline
Advantages & \begin{tabitemize} \item Model utility is preserved \item Achieves the strongest privacy and compliance guarantees \end{tabitemize} & \begin{tabitemize} \item Computationally-expensive \item Difficult to implement in complex models \end{tabitemize}\\ \hline
Weaknesses & \begin{tabitemize} \item Computation- and resource-efficient \item Applicable to complex models \end{tabitemize} & \begin{tabitemize}
\item Parameters may still contain sensitive information
\end{tabitemize}\\ \hline
\end{tabular}
\end{table}

With the growing emphasis on the RTBF, MU has gained significant attention, leading to the development of various techniques for removing data influence from trained machine learning models. Generally, these techniques can be categorized into exact unlearning and approximate unlearning. Exact unlearning~\cite{cao2015forget, bourtoule2021unlearning, brophy2021random, wu2020deltagrad}, also referred to as naive retraining, ensures that the model obtained after processing an unlearning request is provably identical to the model trained with the data targeted for removal. This approach provides the strongest guarantees of privacy and regulatory compliance, making it the "\emph{gold}" standard in MU. However, its computational cost renders it impractical for large-scale models or datasets. In contrast, approximate unlearning~\cite{nguyen2020variational, guo2020certified, golatkar2020eternal, sekhari2021remember, ginart2019forget} relaxes these strict guarantees by eliminating the influence of target data only to a "\emph{quantifiable approximation}". Typically achieved through fine-tuning or modification of trained models, approximate unlearning does not guarantee equivalence to naive retraining but offers greater efficiency in terms of time and computational resources, making it more suitable for practical applications. Table~\ref{tab:exactapproximate} summarizes the differences between exact and approximate unlearning. The inherent trade-off between privacy guarantees and efficiency must therefore be carefully considered when designing MU frameworks. As MU itself is not the central focus of this survey, we refer interested readers to recent comprehensive surveys on MU~\cite{shaik2025mu, liu2025threats, li2025metrics, qu2024insights} for a more detailed understanding of the unlearning approaches in centralized machine learning settings.

Several survey papers have emerged on FUL, motivated by the need to extend the RTBF to FL models trained in distributed and decentralized settings. In~\cite{liu2024survey}, the authors discuss a unified workflow for FUL, emphasizing its role in enabling integration with Machine Learning as a Service (MLaaS). The authors in~\cite{romandini2025survey} place greater emphasis on the guidelines for FUL algorithm design and implementation, categorizing methods according to objectives, techniques and evaluation metrics. However, their work does not adequately highlight the limitations of existing approaches or analyze how FUL techniques address the unique challenges of FL. The study of~\cite{elbedoui2025sok} provides a comprehensive Systematization of Knowledge (SoK) review of FL and FUL within the context of medical image analysis. Their survey is more focused on FL frameworks, presenting FUL methods only superficially, without in-depth technical discussion or insights into their adaptation for healthcare. The authors in~\cite{bharathi2024emerging} highlight emerging challenges in FUL, particularly in incentive mechanisms, environmental sustainability and applications to Large Foundational Models such as LLMs. However, their review remains at a high level, categorizing FUL algorithms without providing technical depth or practical implementation insights. Similarly, the authors in~\cite{zhao2025exploring} examine key challenges, tradeoffs and evaluation metrics in FUL, but adopt a generalized approach, presenting existing frameworks without comparative analysis across surveyed solutions. The authors in~\cite{wang2024privacy} take a more specialized focus, analyzing privacy vulnerabilities in FUL, particularly its susceptibility to membership inference attacks and summarizing applicable defence strategies. Nevertheless, their survey overlooks broader FUL applications and implementation challenges such as data heterogeneity among clients and intermittent participation in FL training rounds. Table~\ref{tab:fusurvey} provides an overview of the selected surveys in FUL, highlighting their key contributions and limitations. 

\begin{table*}[!ht]
\caption{\small Contributions of selected FUL surveys.} 
\label{tab:fusurvey}
\centering
%\begin{tabular}{|l |l| l| l|}
\begin{tabular}{|>{\centering\arraybackslash}m{1cm}|>{\centering\arraybackslash}m{1cm}|>{\centering\arraybackslash}m{0.6cm} | m{6cm}|m{6cm}|}
%{0.75\textwidth}
\hline

\rowcolor{mygray}
\textbf{Subject}   & \textbf{Types} & \textbf{Ref.}  & \multicolumn{1}{c|}{\textbf{Key Contributions}} & \multicolumn{1}{c|}{\textbf{Limitations}} \\ 
\hline

\multirow{19}{*}{\rotatebox[origin=c]{90}{Federated Unlearning}} & \multirow{8}{*}{\rotatebox[origin=c]{90}{Comprehensive Survey}} & \cite{liu2024survey} & \begin{tabitemize} \item Present a unified  FUL workflow for its integration with MLaaS \end{tabitemize} & \begin{tabitemize} \item Lacks a systemic analysis of FUL methods from the perspective of edge implementation challenges \item Discussions on how FUL can be applied to real-world edge applications are largely missing \end{tabitemize}  \\
\cline{3-5}
& & \cite{romandini2025survey} & \begin{tabitemize} \item Discuss the guidelines for the design and implementation of FUL algorithms \item Categorized the existing FUL algorithms based on objectives, techniques and evaluation metrics \end{tabitemize} & \begin{tabitemize} \item Limitations of the FUL algorithms are not highlighted 
 \item Discussions on how the FUL algorithms are designed to mitigate the challenges of FL settings are not included \end{tabitemize}\\
\cline{3-5}
& & \cite{elbedoui2025sok} & \begin{tabitemize} \item Present the Systematization of Knowlwedge (SoK) review of FL and FUL within the context of medical image analysis \end{tabitemize} & \begin{tabitemize} \item More focused on FL than FUL \item Merely presented several FUL algorithms without in-depth technical details \end{tabitemize}\\
\cline{2-5}

& \multirow{6}{*}{\rotatebox[origin=c]{90}{Brief Survey}} & \cite{bharathi2024emerging} & \begin{tabitemize} \item Highlight the key challenges in FUL, particularly in incentive mechanisms designs, environmental sustainability consideration and applicability to Large Foundational Models \end{tabitemize} & \begin{tabitemize}
 \item In-depth technical details of the different FUL frameworks are not provided \item Evaluation metrics of FUL algorithms are not presented \end{tabitemize} \\
\cline{3-5}
& & \cite{wang2024privacy} & \begin{tabitemize} \item Extensive discussion onn potential attacks and defensive measures with great attention on the vulnerability of the FUL systems towards membership inference attacks \end{tabitemize} & \begin{tabitemize} \item Focus mainly on privacy issues in FUL \item Considerations of other implementation challenges of FUL are not included \end{tabitemize}\\
\cline{3-5}
& & \cite{zhao2025exploring} & \begin{tabitemize} \item Highlights the challenges, tradeoffs and evaluation metrics in FUL \item Presents a unified benchmark framework to evaluate FUL algorithms \end{tabitemize} & \begin{tabitemize} \item In-depth technical details of the different FUL frameworks are not provided \item Integration with current FL algorithms is not explored
\end{tabitemize}\\
\hline

\end{tabular}
\end{table*}

Although both FL and MU have received increasing attention, research on FUL remains in its early stages, with no unified overview available to map the current landscape. Considering the rapid progress in this emerging area, a comprehensive survey is both timely and necessary to consolidate existing approaches, classify methodologies, identify limitations and outline promising directions for future research. This survey addresses this gap by systematically reviewing FUL methods, amalyzing their strengths and weaknesses and situating them within the broader context of privacy-preserving and trustworthy machine learning. Accordingly, the main contributions of this survey are as follows:

\begin{itemize}
    \item We provide a detailed overview of the fundamentals of FUL. Specifically, we present a general framework of FL and its role as an enabling technology in mobile edge networks. We then introduce the concept of FUL, outline its major categories and discuss the core challenges in its implementations.
    \item We review and analyze existing FUL schemes that aim to reduce communication overhead in edge networks, highlighting methods that improve scalability and responsiveness under bandwidth and latency constraints.
    \item We examine FUL approaches designed to address vulnerabilities during the FUL process, focusing on techniques that enhance robustness, trust and compliance with privacy requirements.
    \item We discuss FUL framewokrs that explicitly account for device heterogeneity, limited computational resources and energy constraints, emphasizing practical designs suitable for real-world deployment at the mobile edge. 
    \item We provide a systematic analysis of existing FUL methods, identifying their strengths, limitations and interrelationships to guide future research and system design.
    \item We survey recent FUL applications in mobile edge and user-centric services, demonstrating how FUL enables consent-aware, privacy-compliant and adaptive intelligence across distributed environments. discuss and present the applications of FUL in mobile edge computing and user-facing services.
    \item We highlight unresolved issues and promising research avenues, offering a forward-looking perspective on advancing FUL toward scalable, auditable and trustworthy federated intelligence. 
\end{itemize}

%\cite{wang2025incentivizes}

% \begin{table*}[!ht]
% \caption{\small Our Contributions.} 
% \label{tab:contributions}
% \centering
% %\begin{tabular}{|l |l| l| l|}
% \begin{tabular}{| m{8cm}|m{8cm}|}
% %{0.75\textwidth}
% \hline

% \rowcolor{mygray}
%  \multicolumn{1}{c|}{\textbf{Overlapping Contributions}} & \multicolumn{1}{c|}{\textbf{Unique Contributions}} \\ 
% \hline

% \begin{tabitemize} \item able to Reduce the computation latency as the master is able to recover the final result without waiting for the slowest computing node \end{tabitemize} & \begin{tabitemize} \item able to Reduce the computation latency as the master is able to recover the final result without waiting for the slowest computing node \end{tabitemize}  \\
% \hline

% \end{tabular}
%\end{table*}

\subsection{Scope of Survey}

\begin{table}[t]
\scriptsize
  \caption{\small List of common abbreviations used in this paper.}
  \label{tab:table_abb}
  \centering
  \begin{tabularx}{8.7cm}{|Sl|X|}
  \hline
  \rowcolor{mygray}
 \textbf{Abbreviation} & \textbf{Description} \\   \hline

AIGC & Artificial Intelligence Generated Content\\ \hline
CCPA & California Consumer Privacy Act (CCPA)\\ \hline
CNN & Convolutional Neural Networks\\ \hline
FL & Federated Learning\\ \hline
FUL & Federated Unlearning\\ \hline
GDPR & General Data Protection Regulation\\ \hline
HIPAA & Health Insurance Portability and Accountability Act\\ \hline
IoV & Internet-of-Vehicles\\ \hline
KG & Knowledge Graph\\ \hline
LLMs & Large Language Models\\ \hline
MU & Machine Unlearning\\ \hline
RSU & Roadside Units\\ \hline
RTBF & Right to be Forgotten\\ \hline
RTV & Right to Verify\\ \hline
SGD & Stochastic Gradient Descent\\ \hline
TF-IDF & Term Frequency Inverse Document Frequency\\ \hline
V2X & Vehicular-to-Everything\\ \hline
VFL & Vertical Federated Learning\\ \hline

\end{tabularx}
\end{table}

\begin{figure}
    \centering
    \includegraphics[width=\linewidth]{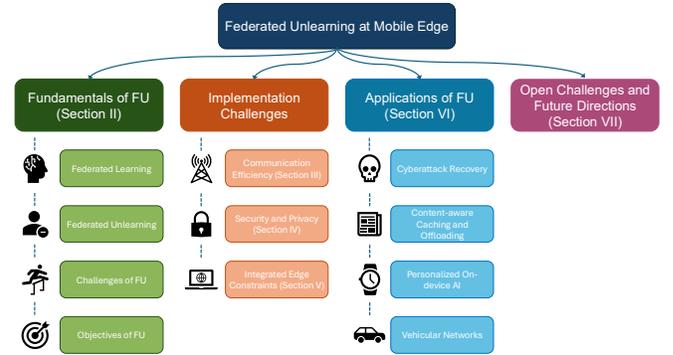}
    \caption{Structure of Federated Unlearning Survey.}
    \label{fig:surveystructure}
\end{figure}

For the reader's convenience, we categorize existing studies on FUL based on the primary challenges they address, namely communication efficiency, security and privacy and edge-specific constraints. This organization allows researchers and practitioners focused on these areas to easily navigate the literature and benefit from our comprehensive analysis, including critical insights into current methodologies, unresolved issues and promising future directions. In Section~\ref{sec:fundamentals}, we present the fundamentals of FUL. Then, we review the FUL frameworks that are proposed to remove the contributions of a specific class, client or sample while addressing the challenges in the federated settings. In particular, Section~\ref{sec:communication-efficiency} presents FUL frameworks that aim to minimize communication costs in the mobile edge networks. Section~\ref{sec:security-and-privacy} discusses the privacy and security approaches in FUL. Section~\ref{sec:integrated-edge-constraints} highlights the FUL frameworks that consider the resource constraints in the mobile edge networks. Section~\ref{sec:applications-ful-mec} and~\ref{sec:open_challenges} discuss the applications and open challenges of FUL respectively. Section~\ref{sec:conclusion} concludes the paper. The structure of the survey is presented in Fig.~\ref{fig:surveystructure}. A list of commonly used abbreviations in this paper is given in Table~\ref{tab:table_abb}.

\section{Fundamentals}
\label{sec:fundamentals}

In this section, we first outline the general FL framework. We then introduce the main categories of FUL frameworks and highlight the key challenges inherent in FL settings that must be addressed when implementing FUL mechanisms. We also discuss the specific objectives of FUL frameworks.

\subsection{Federated Learning}
\label{subsec:federatedlearning}

Given the increasingly stringent privacy regulations and the exponential growth of data generated at the edge, FL has emerged as a promising paradigm for training AI models in mobile edge networks. Unlike centralized training, where models are trained on a centralized server, FL enables training to occur at the edge, close to where data is generated. By avoiding the transmission of raw data, FL allows multiple users to build a shared model collaboratively while maintaining data locality and preserving privacy.

\begin{figure}
    \centering
    \includegraphics[width=\linewidth]{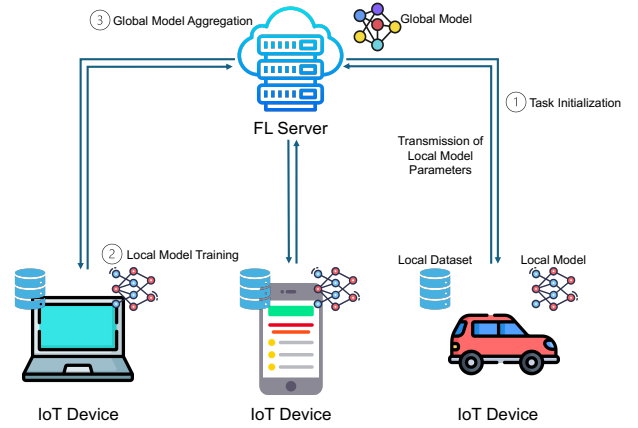}
    \caption{General framework of Federated Learning.}
    \label{fig:federatedlearning}
\end{figure}

In general, an FL system consists of two entities, i.e., model owner (FL server) and the data owners (clients). The goal of the FL server is to train a global FL model with parameters $\boldsymbol{\omega}_G \in \mathbb{R}^d$ using the local datasets of participating clients. Suppose there are $N$ clients in the system, denoted by the set $\mathcal{N} = \{1,\ldots, n, \ldots, N\}$. Each client $n$ possesses a private dataset $D_n = \{(x_n^i,y_n^i)\}_{i=1}^{s_n}$, where $s_n$ denotes the number of data samples owned by client $n$. The FL training process typically involves the following three steps (Fig.~\ref{fig:federatedlearning}):

\begin{itemize}
    \item \emph{Step 1 (Task initialization): }The FL server defines the FL training task, e.g., route recommendation system and medical image analysis, and specifies the hyperparameters of the FL model, e.g., batch size, learning rate and number of local iterations. The initial global model parameters $\boldsymbol{\omega}_G^0$ are then distributed to all participating clients in the network.

    \item \emph{Step 2 (Local model training): }In each iteration $t$, each client $n$ updates its local model by minimizing the local empirical risk function as follows: 
        \begin{equation}
            \boldsymbol{\omega}_n^t = \arg\min_{\boldsymbol{\omega}_G^t \in \mathbb{R}^d}F_n(\boldsymbol{\omega}_G^t) = \frac{1}{s_n}\sum^{s_n}_{i=1}\ell(\boldsymbol{\omega_G^t}, x_n^i, y_n^i),
        \end{equation}
    where $\boldsymbol{\omega}_n^t$ is the updated local model parameters and $\ell(\boldsymbol{\omega_G^t}, x_n^i, y_n^i)$ denotes the prediction loss on the data sample $(x_n^i,y_n^i)$ with the global model parameters $\boldsymbol{\omega}_G^t$. After local training, each client $n$ transmits its updated local model parameters $\boldsymbol{\omega}_n^t$ to the FL server.

    \item \emph{Step 3 (Global model aggregation): }Upon receiving the updated local model parameters from all clients, the FL server aggregates them to update the global FL model by minimizing the global empirical risk as below:
    the updated global model parameters, . The FL server aims to minimize a global empirical risk function defined as below:
    \begin{equation}
        \label{eqn:global}
        \arg\min_{\boldsymbol{\omega}_G^t \in \mathbb{R}^d}F(\boldsymbol{\omega}_G^t) = \sum_{n=1}^{N}p_{n}F_{n}^t(\boldsymbol{\omega}_G^t),
    \end{equation}
    where $p_n = \frac{s_n}{\sum_{n=1}^{N}s_n}$ is the relative weight of client $n$ according to its dataset size. A widely adopted algorithm, \emph{FedAvg}~\cite{mcmahan2017communication}, is used to aggregate the local model parameters of all clients. The FL server minimizes the global empirical risk function in Equation~(\ref{eqn:global}) by computing a weighted average of the local models as follows:
    \begin{equation}
        \boldsymbol{\omega}_G^{t+1} = \sum_{n=1}^{N}p_{n}\boldsymbol{\omega}_n^{t},
    \end{equation}
    where $\boldsymbol{\omega}_G^{t+1}$ is the updated global FL model parameters.
    \end{itemize} 
    For clarity of the readers, we denote the parameters obtained from training on individual client datasets as the \emph{local} model parameters and the model parameters obtained after aggregation as the \emph{global} model parameters. Steps~2 and~3 are repeated iteratively until the model converges to the desired accuracy or the maximum number of training rounds is reached. 

By design, this decentralized training paradigm offers several advantages over centralized approaches~\cite{lim2020federated}. 

\begin{itemize}
    \item \emph{Low Latency: }Since FL models are trained directly on end devices, real-time decision can be made locally without depending on cloud-based inference. This significantly reduces latency compared to scenarios where data must be transmitted to the cloud for processing. Such responsiveness is particularly critical for time-sensitive applications, e.g., autonomous driving systems that reply on rapid actions such as automatic emergency braking.

    \item \emph{Communication-efficient: }FL transmit only model parameters that are typically orders of magnitude smaller than raw datasets to the central server for aggregation. This substantially lowers communication overhead and alleviates stress on backbone networks, making the framework more scalable and resource-efficient.

    \item \emph{Privacy-preserving: }Unlike centralized training, FL does not require users' raw data to be shared with the FL server. By ensuring that data remain local to devices, FL inherently supports privacy preservation and is well aligned with regulatory and societal demands for stronger data protection and ownership rights. 
    
\end{itemize}

\subsection{Federated Unlearning}
\label{subsec:federatedunlearning}

With the advent of increasingly stringent data privacy regulations such as the California Consumer Privacy Act (CCPA) and the European Union's GDPR, clients are granted the RTBF, which entitles them to request the removal of their data contributions from collaboratively trained model. The straightforward approach to achieve this is naive retraining, where the model is retrained from scratch after excluding the data samples that are requested for deletion. However, naive retraining is computationally expensive and often impractical, since if assumes the availability of the same set of clients and resources as in the original training.

Despite its limitations, naive retraining serves as the \emph{gold standard} for FUL as it achieves the same performance as exact machine unlearning. Consequently, FUL can be formally defined as follows: 
\begin{definition}
The objective of an FUL process is to remove the influence of specified data samples from the trained global FL model parameters $\boldsymbol{\omega}_G$, while ensuring that the resulting unlearned model parameters $\tilde{\boldsymbol{\omega}}_G$ perform as closely as possible to the retrained-from-scratch model parameters $\bar{\boldsymbol{\omega}}_G$.
\end{definition}

The FUL approaches are broadly divided into three main categories, namely: class unlearning, client unlearning, and sample unlearning. The differences are illustrated in Fig.~\ref{fig:unlearningvariants}.

\begin{figure}
    \centering
    \includegraphics[width=\linewidth]{unlearningvariants.png}
    \caption{Different types of Federated Unlearning.}
    \label{fig:unlearningvariants}
\end{figure}

\subsubsection{Class Unlearning}For class-level FUL, the objective is to remove all data associated with a specific class label. In particular, a specific class is removed from the global model generalization boundary and the global FL model loses the classification capability on the removed class.

In~\cite{wang2022discriminative}, the authors propose a class-discriminative pruning strategy that quantifies the class relevance of channels using the Term Frequency Inverse Document Frequency (TF-IDF)~\cite{paik2013tfidf}. This approach identifies and removes the most influential channels associated with the target class to be unlearned. While this method achieves performance comparable to retraining from scratch and provides a significant speedup in the unlearning process, its applicability is limited. Specifically, the strategy is tailored to Convolutional Neural Networks (CNNs) and is not easily transferable to other model architectures.

\subsubsection{Client Unlearning}For client-level FUL, it involves the elimination of all data contributed by a particular client. Specifically, the previously learned knowledge from the target client needs to be removed.

A straightforward and classical approach to client-level FUL is the \emph{FedEraser} algorithm proposed in~\cite{liu2021federaser}. \emph{FedEraser} reconstructs the unlearned model by leveraging the historical parameter updates of clients that have been stored at the centralized FL server during the FL training process. This method strikes a balance between storage overhead and reconstruction time. Specifically, retaining more historical updates reduces the number of calibration rounds required, thereby shortening the reconstruction process. The reconstruction process can be sped up by changing the size of the retaining interval or the calibration ratio. Specifically, by having larger retaining interval of the clients' parameter updates in the FL training process, fewer retaining rounds are involved, and thus shorter time is required for the reconstruction of the unlearned FL model. Experimental results demonstrate that \emph{FedEraser} achieves up to a 4$\times$ speedup compared to naive retraining.

It is relatively easy to integrate \emph{FedEraser} into existing FL systems since it requires no modification to the underlying FL architecture or framework. To support client-level FUL, the FL server only needs to retain the model parameter updates from each training iteration. These stored updates can later be calibrated to remove the influence of a specified client upon receiving an unlearning request.

\begin{table*}[t]
\caption{Types of Federated Unlearning.} 
\label{tab:typesunlearning}
\centering
%\begin{tabular}{|l |l| l| l|}
\begin{tabular}{|>{\centering\arraybackslash}m{1cm}|>{\centering\arraybackslash}m{0.6cm} |>{\centering\arraybackslash}m{1.3cm} |m{13.4cm}|}
%{0.75\textwidth}
\hline

\rowcolor{mygray}
\textbf{Types}  & \textbf{Ref.} & \textbf{Methods} & \multicolumn{1}{c|}{\textbf{Key Ideas}} \\ \hline

Class  & \cite{wang2022discriminative} & - & Introduces Term Frequency Inverse Document Frequency to quantize the class discrimination of channels and remove the most relevant channel of the unlearned category \\
\hline

Client & \cite{liu2021federaser} & FedEraser & Calibrate the stored historical model parameters update during FL training process and reconstruct the unlearned model using calibrated updates  \\
\cline{2-4}
  & \cite{wu2022knowledge} & - & Update the global model parameters at each iteration and retrain the unlearned model through knowledge distillation \\
\cline{2-4}
  & \cite{yuan2023recommendation} & FRU & Propose a two-step strategy: (i) small-sized negative sampling and (ii) important-based update selection to efficiently utilize clients' storage space to store historical model updates that are re-calibrated to reconstruct the federated recommendation models \\
\cline{2-4}
  & \cite{halimi2023erase} & - & Formulate the local unlearning problem as a constrained maximization problem by restricting the model parameters to an $\ell_2$-norm ball around a suitable reference model obtained from the average of other clients' model \\
\cline{2-4}
  & \cite{wang2024vertical} & RS2 & Employs RAdam and SGDM in the early and later stage respectively to manage the tradeoffs between model generalization and fast convergence \\
\cline{2-4}
  & \cite{liu2022revfrf} & RevFRF & Ensures that the contributions of the honest revoked clients are not available to the remaining participating clients and the dishonest revoked clients are not able to continue to use the trained FL model \\
\cline{2-4}
  & \cite{su2023asynchronous} & KNOT & Assign the participating clients into different cluster with the aim of minimizing the time required for the retraining process \\
\cline{2-4}
  & \cite{zhang2023fedrecovery} & FedRecovery & Removes a weighted sum of gradient residuals from the global FL model and utilizes the Gaussian mechanism to close the gap between the unlearned and retrained model \\
\cline{2-4}
  & \cite{cao2023fedrecover} & FedRecover & Uses historical information, \emph{Cauchy mean value} theorem and L-BFGS based algorithm to estimate each client's model update after detecting the malicious clients \\
\cline{2-4}
  & \cite{gong2022forgetsvgd} & Forget-SVGD & Represents the variational posterior distribution in a non-parametric fashion through a set of global particles and updates local SVGD by minimizing the local unlearning free energy problem \\
\hline

Sample  & \cite{liu2022rapid} & - & A rapid retraining method based on the diagonal empirical Fisher Information Matrix (FIM) with an adaptive momentum technique \\
\cline{2-4}
& \cite{sheng2024robust} & robustFU & Employs a dynamic conflict sample compensation algorihm and an innovative global reweighting mechanism \\
\cline{2-4}
& \cite{exactfun2023xiong} & Exact-Fun & Introduces Q-FL algorithm to facilitate the exact FUL process that improves efficiency and effectiveness of the FUL framework  \\
\cline{2-4}
& \cite{che2023ffmu} & FFMU & Leverage PCMU to train local machine unlearning model on each device and forumlate the local models as output functions of a Nemytskii operator based on nonlinear functional analysis \\
\cline{2-4}
& \cite{zhu2023kgembedding} & FedLU & In the FL stage, mutual knowledge distillation is used to transfer local knowledge to the global and absorb the global knowledge back. In the FUL stage, retroactive interference and passive decay are used to remove the influence of a specific triplet on the global FL model \\
\cline{2-4}
& \cite{liu2020learntoforget} & Forsaken & Implement memorization elimination through a dummy gradient generator\\
\hline

\end{tabular}
\end{table*}

\subsubsection{Sample Unlearning}For sample-level FUL, previously learned knowledge obtained from part of the targets client's dataset needs to be removed from the global FL model. It involves the deletion of individual data samples without affecting the rest of the client's dataset. This is different from the client-level FUL where the knowledge obtained from the \emph{entire} dataset of the target client needs to be eliminated.

To ensure efficient removal of targeted data contributions, the authors in~\cite{liu2022rapid} propose a rapid retraining method based on the diagonal empirical Fisher Information Matrix (FIM). This proposed method is computationally cheaper than the traditional L-BFGS algorithm~\cite{bollapragada2018lbfgs}. To minimize approximation errors, an adaptive momentum technique is applied to the FIM, thereby improving model stability and accelerating convergence.

While most of the works focus on a specific type of unlearning task, i.e., class, client or sample unlearning, some works address more than one type of unlearning task. For example, the momentum degradation (MoDe) method proposed by the authors in~\cite{zhao2024mode} is capable of both client and class unlearning. MoDe involves two stages: knowledge erasure and memory guidance. In the first step, instead of abruptly deleting target updates, the parameter distributions are progressively updated such that the unlearned model is updated in the right direction. In the second step, memory guidance fine-tunes the model parameters to align unlearned models with retrained counterparts. This dual-task approach significantly reduces execution time compared to naive retraining and strengthens resilience against data poisoning attacks.

\subsection{Challenges of Federated Unlearning}

FUL inherits the distributed and privacy-preserving characteristics of FL, while adding the complex requirement of selectively erasing learned information. Unlike MU, i.e., centralized unlearning, FUL must operate over decentralized, asynchronous and privacy-constrained environments, where model updates, instead of raw data, are exchanged. These conditions introduce several technical challenges that complicate the design of efficient and verifiable unlearning mechanisms. Among them, three core challenges~\cite{wu2022distillation} stand out: the stochastic nature of FL training, the incremental learning dynamics and the limited accessibility of user data. 

% \begin{figure}
%     \centering
%     \includegraphics[width=0.7\linewidth]{figures/unlearning_overview_flow.png}
%     \caption{Challenges of federated unlearning in mobile edge networks. The mind map organizes the design space along communication efficiency, security and privacy, and integrated edge constraints, highlighting issues.}
%     \label{fig:implementationchallenges}
% \end{figure}

\begin{itemize}

    \item \emph{Stochastic Training Process: }FL typically relies on stochastic optimization algorithms such as Stochastic Gradient Descent (SGD) or its variants to aggregate model updates from multiple distributed clients. This stochastic nature introduces randomness into both the parameter updates and the global model trajectory. Consequently, it becomes challenging to precisely trace or isolate the influence of any individual client or data point on the final model. When a deletion request arises, reproducing the exact contribution of the target data is often infeasible due to randomness in batch sampling, local updates and communication delays. Therefore, designing FUL algorithms that can achieve exact or approximate forgetting under stochastic training remains a significant challenge, especially without requiring complete retraining from scratch.
    
    \item \emph{Inremental Learning Process: }FL operates in an inherently incremental or continual learning regime, where new clients and data samples continuously join the FL training process over time. This dynamic environment complicates unlearning as the effects of a specific client's data may propagate across multiple training iterations and indirectly influence other participants' trajectories. Simply removing the historical gradients or parameters associated with the client is insufficient as later model states depend on those updates. Effective FUL mechanisms must therefore model and reverse this temporal dependency while maintaining model stability and accuracy. Balancing continual learning and continual forgetting is particularly challenging in evolving federated environments. 

    \item \emph{Limited Access to Dataset: }In FUL, direct access to raw client data is typically restricted due to privacy constraints, which is the primary motivation for adopting FL in the first place. This limited accessibility makes it impossible to directly retrain or selectively remove data samples. The unlearning process must instead rely on surrogate information such as model updates, parameter differences or influence approximations. However, these indirect signals may not fully capture the statistical impact of the deleted data, leading to incomplete or approximate forgetting. Designing FUL mechanisms that can operate effectively under data unavailability, while still guaranteeing deletion fidelity and privacy preservation, is therefore an essential but difficult problem.
\end{itemize}

\subsection{Objectives of Federated Unlearning}

\begin{figure}
    \centering
    \includegraphics[width=0.8\linewidth]{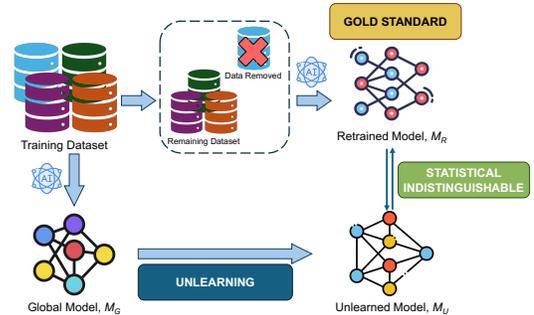}
    \caption{Achieving statistical indistinguishability between the retrained and unlearned models.}
    \label{fig:unlearningretraining}
\end{figure}

The goal of FUL is to enable selective and verifiable removal of clients' data contributions from distributed learning models without compromising performance of privacy. Since FL operates under decentralized and resource-constrained conditions, FUL frameworks must balance multiple, often competing, objectives to remain practical and trustworthy. The following key objectives define the design principles and performance metrics of effective FUL systems.

\begin{enumerate}
    \item \emph{Effectiveness: }Restores the FL model to a state that is statistically equivalent to the one trained from the retraining from scratch method without significant decrease in performance (Fig.~\ref{fig:unlearningretraining}).
    \item \emph{Efficiency: }Achieves forgetting through lightweight corrections or localized updates while minimizing latency and energy consumption such that FUL can be executed quickly on edge devices with limited resources.
    \item \emph{Privacy-preservation: }Operates under the same privacy guarantees as standard FL where clients do not disclose their sensitive private data.
    \item \emph{Certified Removal: }Involves formal proofs or empirical tests that demonstrate that the parameters and outputs of the FL model are independent of the deleted data.
\end{enumerate}

\section{Communication Efficiency}
\label{sec:communication-efficiency}

\begin{table*}[!t]
\centering
\caption{\small Classification of Federated Unlearning methods by Communication Efficiency with Strategies, References, Core Mechanisms, Advantages, and Limitations.}
\label{tab:comm-efficiency}
\setlength{\tabcolsep}{4pt}
\begin{tabular}{|>{\centering\arraybackslash}m{1.3cm}|>{\centering\arraybackslash}m{1.2cm}|>{\centering\arraybackslash}m{2.2cm}|m{4.1cm}|m{4.0cm}|m{4.0cm}|}
\hline
\rowcolor{mygray}
\textbf{Subcategory} & \textbf{Strategy} & \textbf{References} & \textbf{Core Mechanism} & \textbf{Advantages} & \textbf{Limitations} \\
\hline

\textbf{Provable and server-side exactness} &
Exact reconstruction &
\textit{FATS}~\cite{Tao2024Provable}, \textit{F$^2$L$^2$}~\cite{jin2024forgettablefederatedlinearlearning}, \textit{Exact-Fun}~\cite{exactfun2023xiong}, \textit{FFMU}~\cite{che2023ffmu} &
The server stores minimal sufficient statistics or reconstruction operators and computes exact or near exact deletions without recalling clients. &
Communication remains low because deletions happen on the server and new rounds are not required in most cases. Certification of deletion is possible and storage is compact compared with full logs. &
The approach needs stability or linearity assumptions during training and it still requires small server metadata. Utility may drift under heavy heterogeneity and fairness must be monitored. \\
\cline{2-6}

& Stability preserving &
\textit{SIFU}~\cite{fraboni2024sifu}, \textit{Client-Free FU}~\cite{fu2024client} &
Training schedules and anchor metadata are designed to enable server only exact removal from limited aggregates. &
Clients do not need to reconnect for deletion and total traffic is reduced during unlearning. Idle synchronization is avoided. &
Anchors and metadata must be maintained carefully and performance can be sensitive to non IID data. Additional auditing may be needed to ensure fairness. \\
\hline

\textbf{Selective and rollback-based strategies} &
Targeted rollback &
\textit{FastRec-FU}~\cite{zhou2025federated}, \textit{FU-PA}~\cite{zeng2025fu}, \textit{FedEraser}~\cite{liu2021federaser}, \textit{FRU}~\cite{yuan2023recommendation} &
The server reverts the specific contributions of forgotten clients using stored deltas snapshots or discriminant corrections with short recovery. &
Only affected parameters or clients are touched so communication rounds are reduced. Recovery is fast and global restarts are usually not required. &
Accurate logs or statistics are needed and storage can raise privacy questions. Residual influence can persist under non IID data and rare patterns may be impacted. \\
\cline{2-6}

& Perturbation based &
\textit{Goldfish}~\cite{wang2024goldfish}, \textit{NoT}~\cite{khalil2025not}, \textit{PAR}~\cite{pang5357506perturb}, \textit{PGD}~\cite{halimi2023erase}, \textit{Rapid Retraining}~\cite{liu2022rapid} &
Parameter updates apply bounded negation or perturbation and a brief recovery step restores accuracy. &
The method communicates less than full retraining and client participation remains small. Warm starts speed up convergence and reduce traffic. &
Deletion is approximate and requires careful tuning. Small recovery phases may still involve clients and guarantees are weaker than exact methods. \\
\cline{2-6}

& Pruning based &
\textit{CD-Pruning}~\cite{wang2022discriminative} &
Discriminative channels or layers are pruned and a minimal replay with short fine tuning cleans remaining effects. &
Most edits are server side and payloads remain small. The model regains utility quickly with few client wake ups. &
Aggressive pruning can harm accuracy and noisy attribution can bias the result. Replay must be calibrated for stability. \\
\hline

\textbf{Clustering, partitioning, and participation control} &
Clustered unlearning &
\textit{FednP}~\cite{jia2025fednp}, \textit{MUFC}~\cite{pan2023mufc}, \textit{FedCIO}~\cite{qiu2023fedcio} &
Clients are organized into clusters or shards so deletion is localized and aggregation can be performed in one shot when possible. &
Only a bounded subset of clients participates which lowers uplink cost and wall clock time. Exact aggregation is possible in some designs. &
Cluster assignments can drift and bookkeeping adds overhead. Accuracy can suffer if partitions diverge or are imbalanced. \\
\cline{2-6}

& Async and coded &
\textit{KNOT}~\cite{su2023asynchronous}, \textit{Isolated \& Coded Sharding}~\cite{lin2024scalable} &
Asynchronous clustered updates isolate the affected group and coded sharding limits the number of contacted clients and stored states. &
Retraining proceeds faster because other clusters do not wait and fewer clients are activated. Server storage can be reduced through coding. &
Extra coordination is required and coding introduces compute overhead. Poor shard boundaries can reduce accuracy. \\
\cline{2-6}

& Subparameter control &
\textit{FedUMP}~\cite{zhu2024federated}, \textit{FUL via Subparameter}~\cite{yadav2025federated} &
Edits target selected layers or subspaces and only essential clients are scheduled while unaffected parts remain frozen. &
Messages are smaller and the number of rounds is lower. Unaffected components keep their learned knowledge intact. &
Success depends on correct subspace attribution and some leakage can remain outside the edited region. Scheduling must be tuned carefully. \\
\hline

\textbf{Compression, distillation, and acceleration} &
Compressed updates &
\textit{Forget-DSVGD}~\cite{gong2022compressed}, \textit{Forget-SVGD}~\cite{gong2022forgetsvgd}, \textit{Adaptive Clipping + KD}~\cite{xie2024adaptive} &
Gradients and activations are quantized or clipped and lightweight knowledge transfer maintains utility with smaller payloads. &
Uplink bandwidth decreases while accuracy is preserved in many settings. Training and deletion both benefit from compact signals. &
Utility depends on the fidelity of particles or surrogates and stability can degrade under strong heterogeneity. Privacy of compressed signals requires verification. \\
\cline{2-6}

& Distillation based &
\textit{QuickDrop}~\cite{dhasade2024quickdrop}, \textit{FedAU}~\cite{gu2024unlearningduringlearning}, \textit{FUSED}~\cite{zhong2025unlearning}, \textit{BFU-SS}~\cite{wang2023bfu} &
Teacher student transfer and synthetic retain sets reconstruct knowledge with compact exchanges during deletion. &
The number of rounds often drops and updates are small which improves responsiveness at the edge. &
Effectiveness depends on surrogate fidelity and privacy controls and tuning is needed to avoid drift with non IID data. \\
\cline{2-6}

& Subspace and momentum &
\textit{SFU}~\cite{li2023subspace}, \textit{MoDe}~\cite{zhao2024mode}, \textit{FedUHB}~\cite{jiang2024feduhb}, \textit{LTU}~\cite{wang2024learningtounlearn}, \textit{FedU}~\cite{wang2024fedu}, \textit{FedAF}~\cite{li2025class} &
Updates are projected into low dimensional subspaces and accelerated with momentum or heavy ball dynamics for faster recovery. &
Transmission size is reduced and synchronization rounds are fewer which gives fast convergence and lower wall clock time. &
Approximation quality drives final accuracy and results can be sensitive to subspace design and momentum tuning. Surrogate privacy still needs consideration. \\
\hline
\end{tabular}
\end{table*}

Communication efficiency is a critical bottleneck in FUL at the mobile edge, where devices and access points operate under tight bandwidth, energy, and connectivity constraints~\cite{lim2020fmobileedge}. Unlike standard FL, FUL amplifies communication costs because it often requires additional rounds to remove outdated or malicious updates, redistribute corrected parameters, and resynchronize models across heterogeneous and intermittently connected clients. These extra transmissions become impractical on resource limited mobile and IoT networks where links are unstable, uplink capacity is scarce, and power budgets are minimal.

This challenge comes from the need to balance unlearning accuracy with the reduction of communication overhead. Edge devices must send corrective gradients or rollback requests to the server while simultaneously conserving bandwidth and energy. Without efficient communication design, unlearning may stall, violate latency budgets, or even fail to guarantee consistency across distributed models.

\begin{figure}
    \centering
    \includegraphics[width=0.8\linewidth]{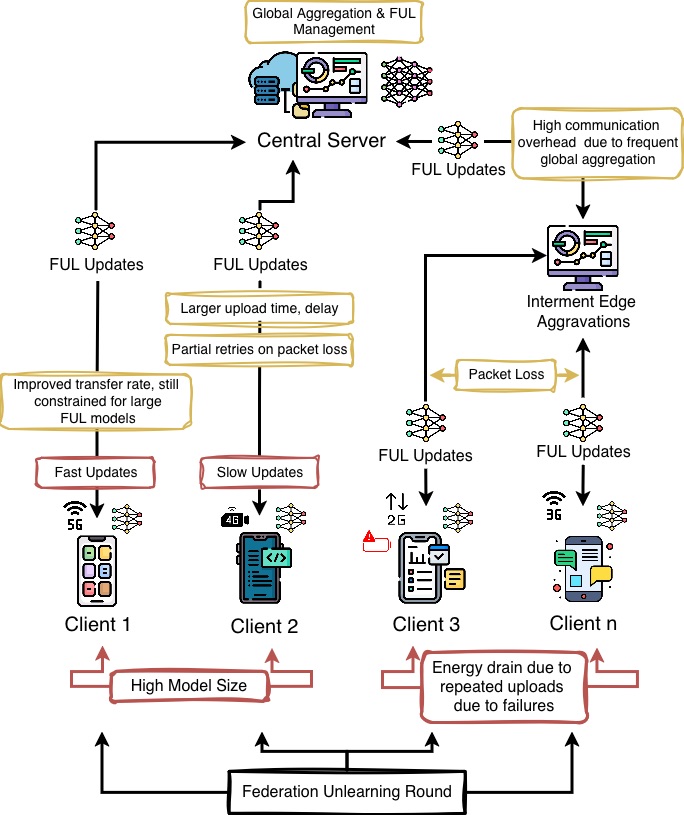}
    \caption{Communication challenges of federated unlearning in mobile edge networks. The figure depicts a hierarchical deployment with mobile and IoT clients connected to edge servers and a cloud FL coordinator, and highlights how unlearning requests introduce extra communication paths on top of standard training rounds. Additional client–edge–cloud exchanges for deletion requests, rollback updates, and model resynchronization stress bandwidth constrained uplinks, amplify the impact of device churn and intermittent connectivity, and make naive retrain-based FUL impractical in real mobile edge environments.}
    \label{fig:challengecommunication}
\end{figure}

Therefore, a communication-efficient FUL is essential to make deletion operations feasible and verifiable in real-world deployments. Figure~\ref{fig:challengecommunication} summarizes communication bottlenecks between mobile clients, edge nodes, and cloud servers under dynamic network conditions. To address these limitations, recent FUL frameworks employ strategies that reduce client–server interactions, compress model exchanges, and localize synchronization without compromising correctness or utility. These solutions can be grouped into four main design families.

\subsection{Provable and server-side exactness}

This family targets \emph{exact or certified} unlearning while shifting nearly all computations to the FL server, so that the clients rarely or never need to be recalled. The key levers are (i) \emph{training-time stability} (e.g., total-variation or quantization stability) that enables analytic deletion later, (ii) \emph{server-only corrections} such as Newton-style or closed-formed updates that transform trained model parameters into counterfactual ones, and (iii) \emph{trajectory or subspace reconstruction} on the FL server that allows unlearning without fresh rounds from edge devices. This design is particularly attractive at the mobile edge, where waking intermittently available, bandwidth-limited clients for post-hoc cleanups is highly disruptive.

Server-side approaches reduce or eliminate client recalls through:
\begin{itemize}
  \item \textit{Zero or few-round unlearning:} Server-only corrections such as Newton step on the linearized objective avoid any new client rounds after an erase request, preventing uplink bursts and idle waits by devices.
  \item \textit{Round compression during training:} Stability-preserving protocols such as TV-stable local-SGD with periodic averaging~\cite{haddadpour2019SGD, reisizadeh2020fedpaq} reduce the number of global synchronizations needed \emph{both} during training and for certified deletions.
  \item \textit{Sufficient statistics in place of raw updates:} Storing compact aggregates, e.g., last-round gradients or quantized updates, avoids recalling clients and limits storage costs since full client re-participation is not required and the FL server needs only small metadata.
  \item \textit{Client-free deletion by design:} Approaches that guarantee client independence make FUL feasible even when target or retained clients are offline or resource-poor, a common reality at the edge.
\end{itemize}

\emph{FATS}~\cite{Tao2024Provable} formalizes exact FUL through total-variation stability and designs a stability-preserving variant of FedAvg that uses local SGD with periodic averaging, achieving statistically indistinguishable deletion with convergence guarantees while cutting the number of communication rounds. It avoids recalling clients at deletion time and keeps the communication footprint low, but it requires training to follow the stability schedule and to maintain its assumptions. Small server-side aggregates may need to be retained, and the tightness of the bounds can depend on the model class and the severity of non-IID data. \emph{\text{F$^2$L$^2$}}~\cite{jin2024forgettablefederatedlinearlearning} proposes forgettable federated linear learning with certified removal by linearizing a pretrained network and executing server-only deletion via a Newton-style correction derived from server-held linearized statistics, yielding certified bounds relative to full retraining. This eliminates additional client communication at deletion time and works well with modern pretrained backbones, although precision and guarantees depend on the quality of the linear approximation and curvature estimates, and the benefit narrows when many layers are fine-tuned.

\emph{Exact-Fun}~\cite{exactfun2023xiong} builds a quantized federated learning pipeline that stabilizes the training trajectory so that exact unlearning can be carried out afterwards under its stated conditions. The construction avoids full retraining and keeps client interaction minimal during deletion, yet requires FL system designers and ML engineers deploying the model to integrate quantization from the start, accept possible utility trade-offs, retain small server-side metadata, and operate within regularity assumptions that underpin the guaranties. \emph{FFMU}~\cite{che2023ffmu} extends certified unlearning to the federated setting with a nonlinear functional perspective, in which clients train local unlearning models and the server aggregates them using a point wise nonlinear aggregation operator that treats the model as a function and applies the same nonlinearity at each coordinate. This design amortizes cost by preparing during training and can reduce the total number of rounds compared with retraining or calibrate baselines, but still requires instrumentation of the training process, preserves some client side computation, and its practical gains depend on task smoothness and the degree of data heterogeneity.

\emph{SIFU}~\cite{fraboni2024sifu} delivers provable client unlearning for FedAvg without adding client-side costs by employing a sequentially informed update rule analyzed in convex and non-convex regimes. It fits standard workflows and reduces the need for extra rounds, although its guarantees rely on smoothness-type conditions, and the realized gains with deep non-convex models depend on tuning and heterogeneity. \emph{Client-Free Federated Unlearning via Training Reconstruction}~\cite{fu2024client} performs deletion entirely on the server by reconstructing a sufficient subspace of the training trajectory so that no clients need to participate at deletion time. This is attractive when devices are offline or cost-constrained and eliminates post-hoc uplink bursts, while shifting the bottleneck to reconstruction quality and server records or storage, with guaranties that are typically approximate and sensitive to drift from the reconstructed subspace.

Compared to other families, provable server-side methods reduce communication in two ways: (1) the \emph{deletion step} is executed entirely on the FL server with zero or few new rounds, and (2) the \emph{training step} is designed to be unlearning-friendly, which reduces synchronization frequency and message sizes. The main trade-offs are up-front training constraints (stability, quantization or linearization) to earn exactness and modest server storage for aggregates and metadata at the fL server. Under harsh edge constraints, e.g., sparse connectivity and strict uplink budgets, these approaches offer the most predictable communication footprint while providing verifiable and reproducible deletions. Nonetheless, fairness and heterogeneity challenge remain open challenges as exactness does not eliminate performance drift under highly non-IID client data, requiring careful auditing in practice.

\subsection{Selective and Rollback-based Strategies}

Selective- and rollback-based unlearning reduces communication overhead by limiting corrections to only the necessary components and reusing previously computed updates. Instead of restarting federation-wide training, the FL server and a small subset of clients apply targeted adjustments that erase the influence of a user, class, or shard, followed by a brief recovery phase to restore utility. In mobile edge deployments where bandwidth is scarce and devices are intermittently available, this approach narrows both the set of participating devices and the number of synchronization rounds. Methods that operate with only the target’s data, or that exchange compact statistics rather than full gradients or model states, further reduce uplink demand and keep battery-constrained devices idle. Empirical results from recent systems show that scoped correction plus short calibration can substantially reduce overall communication while retraining accuracy close to retraining from scratch.

\textit{FedEraser}~\cite{liu2021federaser} demonstrates this principle by replaying server-side stored client updates while omitting the target's contributions, followed by a narrow calibration step. This trades storage and computations of the FL server for reduced communication, avoiding fresh global rounds in bandwidth-limited edge settings where connectivity is sporadic. Similarly, \textit{rapid retraining}~\cite{liu2022rapid} accelerates convergence from pre-deletion model with strong initialization such that the federation communicates far less than a cold restart while reaching comparable accuracy.

In recommendation systems, \textit{FRU}~\cite{yuan2023recommendation} adopts a rollback pattern that logs compact on-device updates, replays and adjusts them centrally to excise the target client's impact and performs a brief recovery. The scope is narrow and the participating set is limited to the relevant parties, which lowers uplink demand and coordination overhead at the edge. \emph{Class-discriminative pruning}~\cite{wang2022discriminative} removes target-class channels fine-tunes briefly to regain utility, minimizing client traffic since most work is orchestrated on the server. Projected conflict gradient ascent~\cite{halimi2023erase} takes a constrained optimization approach to steer model parameters away from the target contribution, then stabilizes with a short recovery. This approach is especially attractive for deletion scenarios where full training data or broad client response is infeasible. 

Other designs focus on integrated rollback strategies. \textit{Goldfish}~\cite{wang2024goldfish} coordinates model choice, loss shaping, and optimization to explicitly bound the extra rounds required for deletions, which in turn caps the total communication cost. \textit{NoT} (Negate to Unlearn)~\cite{khalil2025not} starts from the observation that many FUL methods either store historical updates or require renewed access to the data that should be forgotten, which is often infeasible in practice. NoT proposes a storage-free and data-free alternative in which the server applies a layer-wise weight negation perturbation to the global model and then runs a short fine-tuning phase on retained data. Weight negation is analyzed as a strong yet resilient perturbation that pushes the model away from the optimum while preserving a good position for re-optimization, and theoretically is shown to break inter-layer co-adaptation while maintaining an approximate optimality property. 
Empirically, NoT supports client-wise, class-wise, and instance-wise forgetting without accessing the forgotten data, and the authors report that it matches retraining on retain/forget/test accuracy and membership inference metrics while significantly reducing communication and computation overhead across several datasets and architectures~\cite{khalil2025not}. \textit{Perturb and Recover}~\cite{pang5357506perturb} follows a related pattern by erasing targeted behaviors via carefully designed perturbations followed by a short calibration phase that restores performance on retained data. Across these designs and the two methods above, selective rollback closes the mobile-edge communication gap by concentrating work into narrow server-heavy corrections, avoiding broad client recall, and limiting follow-up to a few focused rounds with small payloads.

\textit{FU-PA}~\cite{zeng2025fu} removes the target client's influence by attenuating only those parameters most associated with the target client using discriminant matrices built from first-order signals and aggregates from retained clients. The correction is computed at the server and followed by a short recovery with the remaining clients, which avoids federation-wide restarts and cuts the number of synchronization rounds. \textit{RFUL}~\cite{zhou2025federated} further extends rollbacks to support \emph{recoverable unlearning}, allowing clients to erase and later reinstate their contributions without retraining from scratch. It combines knowledge unlearning to excise the target client’s impact with a knowledge recovery step that rebuilds performance without retraining from scratch, keeping participation narrow and rounds bounded.

Across the selective and rollback-based designs, a common pattern is to (i) undo the target's signal with server-driven corrections, for example using stored deltas, compact subspace edits, or lightweight model surgery, and (ii) stabilize performance with a short recovery phase. This design minimizes client recall, bounds the number of extra rounds, and delivers predictable communication footprints while preserving model utility, thereby addressing the challenge of bandwidth-constrained, intermittently connected mobile-edge environments. Remaining challenges include the reliability of update logs or statistics for rollback, preventing residual leakage or collateral forgetting under non-IID data distributions, and ensuring fairness when deletions disproportionately affect minority or rare patterns.

\subsection{Clustering, partitioning, and participation control}
\label{sec:cluster-participation}

Clustering, partitioning, and participation control aim to localize the impact of unlearning so that only a bounded subset of clients or parameters is affected and only a few rounds are needed. Client-side clustering limits the ripple of a deletion to the cluster that contains the target, sharding divides the federation into isolated groups that retrain locally, and participation control schedules only the minimum necessary clients while others continue training or remain idle. Recent systems instantiate these ideas with asynchronous clustered aggregation that walls off clusters during retraining, with isolated and coded sharding that reduces both the number of affected clients and the server storage cost, and with privacy-aware clustering that integrates secure aggregation under dynamic participation. Partitioning can also be applied in parameter space through subspace clustering and selective freezing so that only the neurons tied to the target are affected during deletion. These patterns directly suit mobile-edge deployments by shrinking the active set per round and concentrating communication within a narrow corridor of clients or parameters.

\emph{KNOT} introduces asynchronous FU~\cite{su2023asynchronous} with clustered aggregation so that retraining is restricted to the target cluster while other clusters continue their normal training. Its client-to-cluster assignment optimization minimizes wall-clock retraining time and achieves up to 85\% faster retraining than competing baselines in the asynchronous setting. From a networking standpoint, cluster-level restriction significantly reduces uplink demand and airtime per cell because only clients in the affected cluster transmit unlearning updates, while clients in other clusters keep following the standard training schedule. This also limits the number of edge-cloud synchronization events triggered by each deletion, since only the model state of the target cluster needs to be reconciled with the cloud rather than the entire federation. The design relies on asynchronous coordination and on forming clusters that balance retraining speed with data heterogeneity, which increases the complexity of server-side orchestration and state management and may require additional control-plane signaling to maintain cluster membership and progress. Similarly, \emph{FedCIO}~\cite{qiu2023fedcio} follows a related spirit with client clustering, isolated retraining within clusters, and one-shot aggregation of cluster models to achieve exact unlearning. Once the target cluster has completed its isolated retraining, the cloud aggregates a small number of cluster-level models in a single step, which reduces backhaul traffic and avoids repeated global broadcasts. As with KNOT, the communication and latency benefits come with the risk of distribution drift if clusters become highly divergent over time, which can increase the need for occasional re-clustering or rebalancing and introduce extra control messages on the edge and backhaul links.

\emph{Isolated and coded sharding}~\cite{lin2024scalable} divides clients into isolated shards so that deletions touch only the target shard. By compressing cross-shard parameters with coded computing, it reduces both retraining time and server storage, improves accuracy and keeps membership inference risk in check. Communication overhead at the edge decreases because the number of both active clients and model states exchanged per correction are reduced. The tradeoff is sensitive to shard boundaries and coding overhead, which may harm accuracy if strong cross-client correlations are split. \emph{FednP}~\cite{jia2025fednp} and \emph{FedUMP}~\cite{zhu2024federated} both confine deletions to specific partitions and reconcile updates with lightweight aggregation. Theses designs shorten wall-clock unlearning, reduce wake-ups and keep uplink demand low, though they require acurate partition attribution and careful bookkeeping.
%\emph{FednP}~\cite{jia2025fednp} proposes FUL with multiple client set partitions and confines deletion and recovery to only the partitions that contain the target contribution, which shortens wall-clock time and limits broadcast scope so the number of active clients and synchronization rounds falls under mobile edge conditions. \emph{FedUMP}~\cite{zhu2024federated} builds on client partitioning to accelerate unlearning by operating only within partitions that intersect the target set and then reconciling with a light aggregation step, which reduces wake-ups and uplink traffic while asking for accurate partition attribution and balanced bookkeeping.

Beyond client grouping, \emph{MUFC}~\cite{pan2023mufc} addresses the unsupervised setting by creating a secure federated clustering pipeline that supports initialization and aggregation for efficient removal of client or data sample. By limiting participation to the affected cluster, it reduces the amount of communication required to restore model utility. Careful initialization and secure aggregation are important to maintain both privacy and efficiency. This work also shows that the FUL mechanism can be simple and provable in this regime. Dynamic user participation work~\cite{liu2024guaranteeing} complements this by combining clustering-based FUL with secure aggregation, ensuring privacy even when users drop out or are removed. Besides, by showing that unlearning can be restricted to the target cluster while other clusters continue, bandwidth demands are kept modest at the edge.

Partitioning can also occur in parameter space. In~\cite{yadav2025federated}, parameter space partitioning is applied to weights and neurons rather than to clients, so that classes or users mapped to specific subspaces of the model can be neutralized by freezing or updating only those subspaces. This reduces communication during deletion because the server coordinates compact corrections in a narrow region of the model, and recovery is short, since the rest of the network remains intact. The approach retains accuracy on unaffected classes while driving the target class to near zero, demonstrating that narrow interventions are possible without federation-wide retraining. Its practical cost is integration during training to learn stable partitions and sensitivity to how well the parameter clustering aligns with the task, which may require tuning and a small amount of auxiliary metadata.

Across these methods, the shared recipe is to confine unlearning to a small scope of clients or parameters and minimize the duration of their participation. and reduce the payloads exchanged during unlearning. Clustered aggregation and isolation ensure that only a fraction of clients exchange updates while the rest of the federation continues normally. Sharding and one-shot aggregation reduce both the number of rounds and the payload size. Secure clustering based on aggregation preserves privacy intact under dropouts, and parameter-space-partitioning turns deletions into compact model edits. Collectively, by reducing the number of synchronization rounds and size of data packets, these strategies deliver predictable bandwidth and energy footprints under mobile-edge constraints while maintaining accuracy close to counterfactual retraining.

\subsection{Compression, distillation, and acceleration}
\label{sec:compression-distillation-acceleration}

Compression, distillation, and acceleration aim to reduce both the number of rounds and the bytes per round in FUL by shrinking update payloads, transferring knowledge through compact surrogates and accelerating convergence. Compression reduces payload sizes through quantization, sparsification, and low rank updates, distillation replaces heavy data or full gradients with compact synthetic sets or logits, and acceleration shortens the unlearning horizon using momentum-based dynamics or by embedding forgetting into training itself. For mobile edge deployments where uplinks are scarce, these levers directly reduce device wake-ups and uplink demand because the federation exchanges fewer and smaller updates while the server performs heavier computations.

Knowledge transfer methods reduce communication by substituting raw data or full model deltas with compact surrogates. \textit{Adaptive clipping with knowledge distillation}~\cite{xie2024adaptive} calibrates client updates and transfers knowledge through soft targets to maintain accuracy with small messages. \textit{QuickDrop}~\cite{dhasade2024quickdrop} distills tiny synthetic sets that  replace full local data sets during forgetting, reducing both local compute and cross-device traffic and allowing short recovery with competitive precision. Learning to unlearn~\cite{wang2024learningtounlearn} frames forgetting as a meta-learning task that aligns remembering and erasing gradients, reducing wasted calibration steps and maintaining accuracy with limited exchanges. Knowledge overwriting with sparse adapters in \textit{FUSED}~\cite{zhong2025unlearning} attaches lightweight modules that absorb and reverse target signals while freezing most base model parameters, which bounds communication and enables lightweight rollbacks when policies or requests change. These methods concentrate updates into compact artifact and reduce the length of recovery phase, though they hinge on the quality of distilled data or auxiliary logits and must address privacy risks when sharing synthetic surrogates.

Bayesian and particle-based designs deliver calibrated forgetting with communication savings. \textit{Forget-SVGD}~\cite{gong2022forgetsvgd} removes target's influence by editing or reweighting a particle set that approximates the posterior, allowing the server to steer particles without full retraining or broad client recall. Compressed particle-based federated Bayesian learning and unlearning~\cite{gong2022compressed} applies quantization and sparsification across particles, reducing link usage while preserving the benefits of Bayesian calibration. \textit{BFU-SS}~\cite{wang2023bfu} introduces parameter self-sharing to balance erasure and retention, which stabilizes accuracy during forgetting with modest additional signals rather than full trajectories. These methods reduce communication overhead since they send only a few particles or compressed summaries, through they often require server-side bookkeeping and can raise per-iteration computation cost compared to purely deterministic baselines.

Subspace and momentum-based accelerators localize edits to narrow model regions, keeping communication at its minimum. \textit{SFU}~\cite{li2023subspace} constructs an orthogonal subspace from retained clients and performs gradient ascent in that subspace to erase the target, eliminating the need for full-history storage and requiring only small representation statistics plus short calibration. Momentum degradation in \textit{MoDe}~\cite{zhao2024mode} erases implicit memory by attenuating accumulated momentum and then guides the model back to a stable region, which can shorten recovery while keeping updates compact. During the process, an auxiliary model is updated and this adds computation costs and potentially extra exchanges under non-IID data distribution. \textit{FedU}~\cite{wang2024fedu} executes the influence approximation only at the client requesting deletion, narrowing participation since only clients that request for unlearning perform the heavy work while others continue as usual. This also eases compliance when many devices are offline. \textit{FedAF}~\cite{li2025class} uses neuro-inspired active forgetting mechanism that generates teacher-student memories to overwrite class-wise signals. It limits the number of extra rounds required and preserves benign performance, but requires careful calibration of memory generation to avoid collateral damage.

Explicit accelerators shorten unlearning by speeding up optimization or embedding forgetting into training. \textit{FedUHB}~\cite{jiang2024feduhb} employs Polyak heavy-ball dynamics with a dynamic stopping rule to retrace toward counterfactual solution in fewer steps, which reduces the number of global synchronizations rounds and therefore uplink usage at the edge. \textit{FedAU}~\cite{gu2024unlearningduringlearning} integrates a lightweight auxiliary unlearning module into standard FL training so that deletions can be executed during training without a separate expensive phase, thereby reducing wall-clock latency between request and compliance and keeping communication predictable. Together with the distillation and subspace families, these accelerators illustrate that latency can be reduced without sacrificing retained accuracy when stability and stopping are designed with unlearning in mind.

Across compression, distillation, and acceleration, the shared principle is to minimize payloads and shorten recovery by replacing heavy federation-wide retraining with compact updates, synthetic surrogates, or accelerated trajectories. These strategies directly reduce communication footprints in mobile-edge conditions while maintaining accuracy close to retraining from scratch, though they introduce tradeoffs in privacy, bookkeeping, and stability under non-IID settings.

\subsection{Summary and Lessons Learned}
\label{sec:summary-lessons}

In this section, we consolidate the insights gained from the surveyed works on communication-efficient FUL at the mobile edge. The main observations, grounded in evidence from the cited systems and frameworks, are summarized in Table~\ref{tab:comm-efficiency} and discussed below.

\begin{itemize}
  \item \textbf{Provable server-side exactness} approaches such as \emph{FATS}~\cite{Tao2024Provable}, \emph{F$^2$L$^2$}~\cite{jin2024forgettablefederatedlinearlearning}, and \emph{Exact-Fun}~\cite{exactfun2023xiong} demonstrate that communication can be reduced to zero or a few additional rounds when training is stability preserving or linearizable. These works confirm that analytic deletion on the server, achieved through total-variation or Newton-style corrections, maintains certified equivalence to retraining. However, they also reveal that such guarantees come at the cost of storing small but persistent aggregates on the server and adhering to training schedules that may constrain model flexibility. Thus, \emph{provability trades adaptivity for predictability}, making these designs best suited for regulated or resource-constrained edge deployments where verification outweighs agility.

  \item \textbf{Selective and rollback-based strategies} (e.g. \emph{FedEraser}~\cite{liu2021federaser}, \emph{FRU}~\cite{yuan2023recommendation}, \emph{NoT}~\cite{khalil2025not}, and \emph{FU-PA}~\cite{zeng2025fu}) empirically validate that replaying or inverting stored deltas can achieve retention of utility of 90\% while reducing communication by up to an order of magnitude compared to retraining. These methods show that targeted deletions can be orchestrated by the FL server with narrow client involvement, but also expose practical challenges: maintaining reliable logs under privacy budgets, and ensuring that rollback does not amplify bias under non-IID data. The lesson is that \emph{rollback precision governs efficiency}: finer-grained logs enable cheaper deletions, but increase storage and privacy risk.

  \item \textbf{Clustering, partitioning, and participation control designs} such as \emph{KNOT}~\cite{su2023asynchronous}, \emph{FedCIO}~\cite{qiu2023fedcio}, and \emph{Isolated and coded sharding}~\cite{lin2024scalable} empirically demonstrate that the binding of the active client set sharply cuts the uplink demand and wall clock time - e.g., \emph{KNOT} achieves up to 85\% faster retraining by restricting updates to the affected cluster. These results confirm that structured participation delivers scalability without full synchronization, especially under heterogeneous or intermittent connectivity. However, experiments also show that cluster imbalance or drift across partitions can degrade accuracy and fairness. Hence, \emph{scoping communication works only as well as cluster coherence is maintained}, making adaptive regrouping and monitoring essential at the edge.

  \item \textbf{Compression, distillation, and acceleration} methods, such as \emph{QuickDrop}~\cite{dhasade2024quickdrop}, \emph{Forget-SVGD}~\cite{gong2022forgetsvgd}, \emph{SFU}~\cite{li2023subspace}, and \emph{FedAU}~\cite{gu2024unlearningduringlearning} provide consistent evidence that distillation and synthetic data transfer can achieve comparable recovery in 20–40\% fewer rounds. Their findings collectively indicate that smaller update payloads and accelerated dynamics (e.g., Polyak momentum in \emph{FedUHB}~\cite{jiang2024feduhb}) reduce both uplink and downlink load. However, these gains rely on the fidelity of distilled surrogates and stability of the approximations under non-IID settings. The key takeaway is that \emph{compression saves bandwidth but shifts the bottleneck to representation fidelity and privacy preservation}.
\end{itemize}

Across all families, a consistent pattern emerges. Communication efficiency at the mobile edge is achievable only when \emph{burden of unlearning} is reallocated from many clients to the FL server, from raw data to sufficient statistics, or from direct retraining to compact surrogates. Empirical studies confirm that each strategy lowers communication in different ways through fewer rounds, smaller payloads, or narrower participation, while introducing trade-offs in provability, fairness, or privacy. Therefore, system designers must align their choice of family with deployment priorities such as \emph{certified guarantees}, \emph{fast responsiveness}, or \emph{scalable participation}. 

Ultimately, the lesson learned is that no single method dominates across all dimensions. Rather, hybrid designs that combine \emph{rollback-style local corrections with compressed or distilled updates} appear to be the most promising for the next generation of edge-aware federated unlearning systems.

\section{Security \& Privacy}
\label{sec:security-and-privacy}

\begin{figure}
    \centering
    \includegraphics[width=0.8\linewidth]{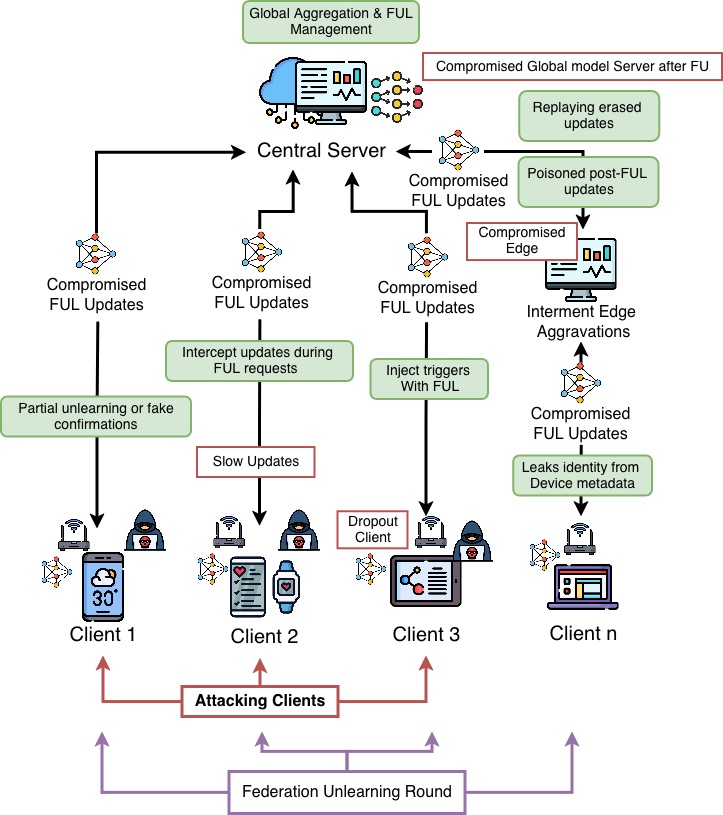}
    \caption{Security Challenges of Federated Unlearning in Mobile Edge.}
    \label{fig:challengesecurity}
\end{figure}

\begin{table*}[!t]
\centering
\caption{\small Classification of Federated Unlearning methods by Security \& Privacy with Strategies, References, Core Mechanisms, Advantages, and Limitations.}
\label{tab:security-privacy}
\setlength{\tabcolsep}{4pt}
\begin{tabular}{|>{\centering\arraybackslash}m{1.7cm}|>{\centering\arraybackslash}m{1.4cm}|>{\centering\arraybackslash}m{2.4cm}|m{3.4cm}|m{3.6cm}|m{3.6cm}|}
\hline
\rowcolor{mygray}
\textbf{Subcategory} & \textbf{Strategy} & \textbf{References} & \textbf{Core Mechanism} & \textbf{Advantages} & \textbf{Limitations} \\
\hline

\textbf{Poisoning and backdoor defense} &
Detection guided &
\textit{FedBT}~\cite{Wang2025FedBT}, \textit{UnlearnGuard}~\cite{wang2025poisoning}, \textit{robustFU}~\cite{sheng2024robust} &
Anomaly scoring and gradient inspection identify malicious clients and poisoning aware objectives filter or down weight suspicious updates before unlearning. &
The system improves robustness because harmful updates are contained early and benign knowledge is preserved during deletion. &
Adaptive attackers may evade detectors and stricter filtering can remove useful signals and reduce accuracy. \\
\cline{2-6}

& Rollback based &
\textit{FedRemover}~\cite{yuan2024toward}, \textit{Get Rid of Your Trail}~\cite{alam2024trail} &
Auditable rollback subtracts the effect of corrupted clients using consistency checks and footprint removal metrics without accessing raw data. &
Recovery quality approaches retraining while maintaining an auditable erasure trail. &
Accurate logs or statistics are required and incomplete footprints can leave residual influence. \\
\cline{2-6}

& Sanitization based &
\textit{FAST}~\cite{guo2023fast}, \textit{Backdoor KD}~\cite{wu2022knowledge} &
Server side sanitization suppresses trigger features and distills safe teacher knowledge to remove adversarial patterns. &
Sanitization is fast and resource friendly and accuracy is preserved on clean data. &
Strong or adaptive triggers can persist and distilled teachers may carry hidden biases. \\
\cline{2-6}

& System level isolation &
\textit{DT-FU}~\cite{daluwatta2024dt-fu}, \textit{UaaS-SFL}~\cite{daluwatta2024uaas} &
Digital twins or unlearning as a service isolate sensitive steps in controlled environments and validate deletions before reconciliation. &
The attack surface is reduced and auditability improves through isolated validation. &
Operational complexity increases and twin fidelity must match real deployments. \\
\hline

\textbf{Certified and verifiable unlearning} &
Client and feature certification &
\textit{Starfish}~\cite{liu2025certifiedclientremoval}, \textit{CFRU}~\cite{huynh2025certified}, \textit{CerFeaUn}~\cite{wang2025-feaun} &
Certificates bound influence after deletion using reconstruction or rollback guarantees and feature level perturbation tests. &
Users receive strong guarantees that deletions are effective and auditors gain measurable evidence. &
Assumptions can restrict model classes and guarantees may be conservative and reduce utility. \\
\cline{2-6}

& Cross party verification &
\textit{Vertical Backdoor Certification}~\cite{han2025verticalbackdoor} &
Inter party certification validates removal of backdoor triggers in vertical federated pipelines across organizations. &
Cross organization assurance increases trust and detects covert dependencies. &
Protocols add coordination overhead and data silo differences complicate tests. \\
\cline{2-6}

& Protocol level auditability &
\textit{VeriFi}~\cite{gao2024verifi} &
Challenge response markers allow participants to confirm deletion without exposing parameters or peer data. &
Participants can verify their own erasure and the system gains tamper evident logs. &
Markers must be carefully designed and verification traffic adds communication cost. \\
\hline

\textbf{Privacy preservation and inference resistance} &
Attribute and instance erasure &
\textit{Aegis}~\cite{wu2025aegis}, \textit{ConFUSE}~\cite{meerza2024confuse} &
Information theoretic or confusion objectives remove sensitive attributes or instances from embeddings to lower inference risks. &
Membership and attribute inference risks are reduced and privacy improves during and after unlearning. &
Erasure can reduce downstream utility and objectives require careful tuning on heterogeneous data. \\
\cline{2-6}

& Cross user mitigation &
\textit{CUFRU}~\cite{li2025cross-user}, \textit{EG-FedUnlearn \& OFU-Ontology}~\cite{ghannam2025ontology-guided} &
Gradient blending temporal decay and ontology alignment suppress leakage that propagates across related users or features. &
Linked entities are cleaned consistently and residual correlations are reduced. &
Ontology curation adds overhead and misalignment can remove useful shared patterns. \\
\cline{2-6}

& DP and secure aggregation &
\textit{FedRecovery}~\cite{zhang2023fedrecovery}, \textit{Dynamic Participation (SecAgg+)}~\cite{liu2024guaranteeing} &
Differential privacy noise and secure aggregation under dynamic participation hide individual updates while supporting retrain equivalent targets. &
Privacy guarantees hold under formal budgets and participation churn is tolerated. &
Noise budgets can impact accuracy and secure aggregation increases protocol complexity. \\
\cline{2-6}

& Deniability and revocation &
\textit{Plausible Deniability FU}~\cite{varshney2025plausible}, \textit{Strategic Data Revocation}~\cite{ding2024strategic} &
Formal deniability for deletion requests and game theoretic strategies guide when users revoke data to balance risk and utility. &
Users gain clear privacy options and systems can plan for rational revocation behavior. &
Deniability can conflict with audit needs and equilibrium strategies may not reflect real user incentives. \\
\hline

\textbf{Cryptographic and blockchain protocols} &
Blockchain anchored audit &
\textit{BlockFUL}~\cite{liu2025blockful}, \textit{BC-MU}~\cite{zuo2025trustworthy} &
Immutable or redactable ledgers record deletion events and model lineage to provide tamper resistant unlearning logs. &
Traceability improves across releases and provenance becomes verifiable. &
On chain costs and latency can grow and privacy must be preserved for sensitive metadata. \\
\cline{2-6}

& Encrypted computation &
\textit{RevFRF}~\cite{liu2022revfrf} &
Homomorphic encryption and multiparty computation enable secure model updates and revocation for tree based FL models. &
Sensitive data remains confidential and deletions are executed without raw data exposure. &
Cryptographic computation increases latency and deployment is complex on resource constrained devices. \\
\cline{2-6}

& Decentralized coordination &
\textit{Decentralized Unlearning (AIGC)}~\cite{lin2024decentralized} &
Coded computing blockchain coordination and differential privacy support unlearning in multi owner generative pipelines. &
End to end traceability improves and coordination across owners becomes feasible. &
System integration is challenging and coded execution can introduce overhead in practice. \\
\hline
\end{tabular}
\end{table*}

Security and privacy have become fundamental concerns in FU, especially at the mobile edge where large numbers of diverse and intermittently connected devices increase both the attack surface and the risk of information leakage. Unlike standard FL, FUL must not only protect data during training but also guarantee that previously learned information is fully and permanently removed. This additional step introduces new vulnerabilities, as deletion requests, correction signals, and intermediate model states can all expose sensitive details or be manipulated by adversaries.

These issues arise largely because FUL operates in distributed and often asynchronous environments where devices differ in capability, connectivity, and trustworthiness. Limited oversight and frequent client turnover make it difficult to verify whether a deletion has been applied correctly or whether adversarial updates have persisted. Without adequate protection, such weaknesses can result in incomplete erasure, residual backdoors, or inference attacks that undermine the integrity and credibility of unlearning.

\begin{figure}
    \centering
    \includegraphics[width=0.8\linewidth]{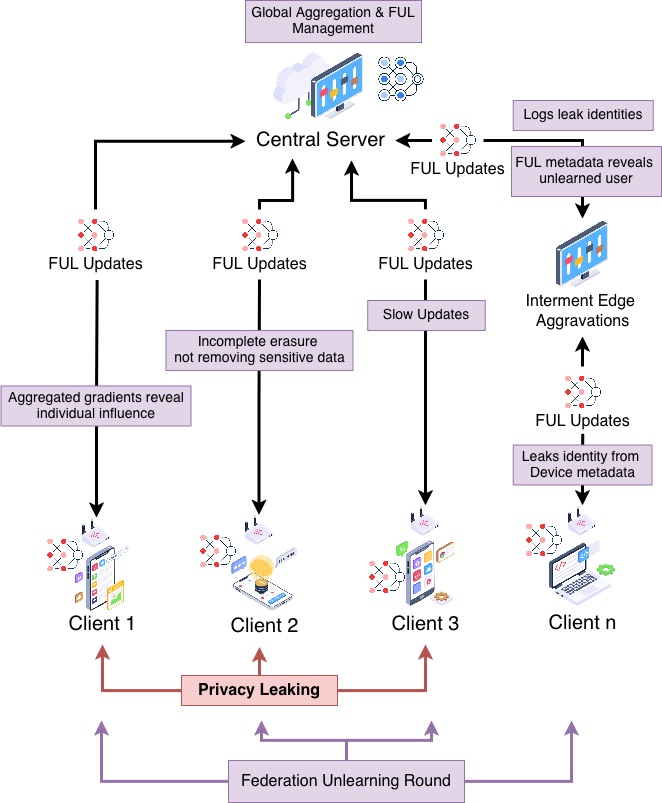}
    \caption{Privacy Challenges of Federated Unlearning in Mobile Edge.}
    \label{fig:challengeprivacy}
\end{figure}

Addressing these challenges is crucial for building trustworthy FUL systems that preserve model integrity, user confidentiality, and regulatory compliance. Figure~\ref{fig:challengesecurity} illustrates common security threats across the communication and aggregation pipelines, while Figure~\ref{fig:challengeprivacy} highlights privacy risks such as data leakage, inference, and partial deletion. To counter these problems, recent studies propose four complementary directions. (i) Poisoning and backdoor defenses that detect and neutralize harmful updates early, (ii) Certified and verifiable unlearning methods that provide auditable evidence of correct deletion~\cite{ginart2019forget}, (iii) Privacy-preserving and inference-resistant techniques that reduce what an adversary can infer from model signals, and (iv) Cryptographic and blockchain-based protocols that harden communication and enable tamper-proof auditing.

\subsection{Poisoning and backdoor defense for security and privacy}
\label{subsec:security-poisoning}

Poisoning and backdoor defense in mobile edge FUL focuses on safeguarding model integrity and user privacy under adversarial behavior. The central goal is to detect and remove malicious influence while preserving benign knowledge and to strengthen the unlearning workflow itself against manipulation. Designs emphasize trustworthy detection, auditable removal, and containment on resource-limited devices where clients churn and data is sensitive.

Detection-guided defenses identify suspicious participants and anomalous updates so that only hostile influence is removed while the rest of the federation remains protected. \textit{FedBT}~\cite{Wang2025FedBT} applies bad-teacher distillation tailored for IoT to steer the student away from poisoned features while retaining benign capacity, improving resilience during and after unlearning. \textit{UnlearnGuard}~\cite{wang2025poisoning} secures the deletion workflow by estimating client updates during unlearning and filtering those that deviate from a clean trajectory so that corrections converge toward a retrain-equivalent baseline. \textit{robustFU}~\cite{sheng2024robust} generates conflict samples and reweights updates to resist adversarial unlearning requests, improving robustness and fairness for remaining clients. These approaches depend on reliable estimators or references and require careful calibration so that false positives do not penalize benign users.

Rollback and subtraction-based defenses offer auditable removal of malicious clients while limiting exposure of sensitive data. \textit{FedRemover}~\cite{yuan2024toward} detects attackers via performance-consistency signals and then executes unlearning to reconstruct a clean model within few rounds, providing a clear removal trail without accessing raw user data. \textit{Get Rid of Your Trail}~\cite{alam2024trail} formalizes footprint stripping from the training record and applies compact corrections so that the final model approximates a clean retrain, while reducing retained metadata that could leak information about the target. These designs improve privacy by avoiding broad data access yet they rely on well-managed logging or telemetry policies that must be minimized and rotated.

Sanitization with compact corrections restores integrity when clean data is scarce or devices are offline. \textit{FAST}~\cite{guo2023fast} performs FUL at the server to identify and eliminate malicious terminals and then quickly reconcile the model to a safe state, which shortens exposure windows under attack. \textit{Backdoor KD}~\cite{wu2022knowledge} subtracts historical attacker influence and then distills benign behavior from the original global model using unlabeled proxy data, thereby purging backdoor signatures while retaining task accuracy. These methods can operate without on-device raw data, but they depend on the availability and quality of safe teachers or small public proxies, raising concerns about auxiliary leakage through transfer.

System-level architectures further isolate sensitive operations and improve privacy guarantees at the edge. \textit{DT-FU}~\cite{daluwatta2024dt-fu} employs a digital-twin layer in vehicular and mobile settings to orchestrate and validate unlearning in the twin space with only lightweight reconciliation at the server. This reduces the on-device attack surface and supports post-hoc inspection. \textit{UaaS-SFL}~\cite{daluwatta2024uaas} realizes unlearning as a service for split FL, running sensitive steps in controlled edge or server environments while devices transmit only intermediate features, which simplifies access control and policy enforcement in regulated domains. These architectures require strong isolation and attestation for the twin or service runtime.

Across these defenses the unifying pattern is to precisely locate hostile influence to protect benign users, provide verifiable removal paths and confine sensitive steps to controlled environments. Detection-guided aggregation reduces the chance of adversarial persistence, rollback and subtraction enable auditable erasure without exposing raw data, sanitization with compact corrections restores clean behavior when on-device access is constrained, and twin- or service-oriented architectures confine risk to hardened components. Together, these defenses align mobile-edge FUL with security and privacy requirements by ensuring model integrity, minimizing user exposure and enabling accountability through inspectable procedures.

\subsection{Certified and verifiable unlearning}
\label{subsec:certified-verifiable}

Certified and verifiable unlearning turns deletion into a security and privacy guarantee for mobile-edge FL systems. The aim is to produce evidence that specific data, clients, or features no longer influence the model while revealing nothing about the underlying data. Certificates may take the form of statistical indistinguishability bounds, formal proofs of removal or verifiable procedures that third parties can check. These mechanisms support accountability auditing and regulatory compliance in sensitive edge settings.

Client-focused certification demonstrates that a named participant has been removed and that the resulting model is close to a clean retrain. \textit{Starfish}~\cite{liu2025certifiedclientremoval} conducts unlearning under two non-colluding servers using two-party computation with secret-shared historical information and provides a theoretical bound that certifies closeness to a retraining-from-scratch baseline without exposing raw updates. \textit{CFRU}~\cite{huynh2025certified} targets federated recommendation and achieves certified client removal by rolling back and eliminating the target client’s historical updates while sampling significant rounds and compensating bias via Lipschitz based estimation, enabling operators to assess whether additional cleanup is needed while preserving privacy.

Feature-focused and vertical certification ensures that specific attributes or channels have been erased. This fits cross-party settings where features are split across organizations. \textit{CerFeaUn}~\cite{wang2025-feaun} introduces certified feature unlearning via feature perturbation and defines the residual effect through parameter differences to bound remaining influence of the forgotten attributes so that the auditors can verify compliance without access to private raw inputs. \textit{Vertical backdoor certification}~\cite{han2025verticalbackdoor} applies backdoor certification tests in VFL to verify that trigger-carrying feature paths have been neutralized while preserving benign functionality across organizations, providing assurance against covert shortcuts in multi-holder pipelines.

\begin{figure}
    \centering
    \includegraphics[width=1\linewidth]{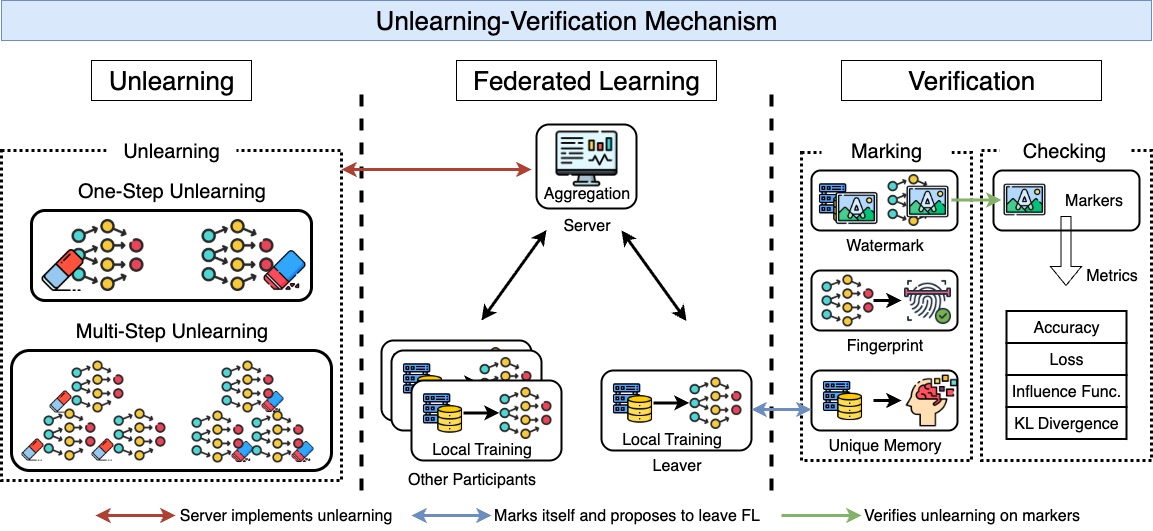}
    \caption{High Level Architecture of \textit{VeriFi}.}
    \label{fig:verifi}
\end{figure}

Protocol-level verifiability builds end-to-end guarantees that the unlearning workflow itself was executed correctly. \textit{VeriFi}~\cite{gao2024verifi} grants the leaving participant the right to verify their removal by embedding markers and later testing the model’s responses over subsequent rounds, enabling verifiers to confirm deletion without learning anything about other clients. This strengthens privacy at the edge and deters manipulation of the removal phase.

Across client, feature and protocol layers, these methods align with security and privacy goals in mobile-edge FUL by replacing trust with proof. Client certificates demonstrate that participants no longer shape the model feature, certificates ensure that targeted attributes and hidden triggers are erased and protocol-level verification makes the entire workflow auditable without exposing data. The shared outcome is stronger integrity, minimal disclosure and clear evidence of compliance that can be checked by external stakeholders.

\subsection{Privacy preservation and inference resistance}
\label{sec:privacy-inference}

Privacy preservation and inference resistance in mobile-edge FUL aim to ensure that the act of deletion does not reveal \emph{who} was removed or \emph{what} was removed and that the resulting model does not leak sensitive membership or attribute information. Methods reshape signals during erase and recovery so that exposures stay within strict bounds, create outcomes that are indistinguishable to adversaries within formal privacy models, align representations to suppress identifiers and protect protocol-level traces such that individual contributions remain hidden even under intermittent participation.

Attribute and instance erasure methods suppress inference risk by directly removing sensitive signals. \textit{Aegis}~\cite{wu2025aegis} performs post-training attribute unlearning in federated recommenders by erasing sensitive attributes from user embeddings using an information-theoretic objective that lowers attribute inference risk while preserving utility for benign behavior. \textit{ConFUSE}~\cite{meerza2024confuse} degrades memorized evidence of the target with confusion-driven updates guided by saliency, enabling instance- and feature-level forgetting without relying on teacher-student transfer, which reduces collateral forgetting when clean data is scarce. \textit{Plausible Deniability FU}~\cite{varshney2025plausible} provides proof of deniability for aggregated updates so that the server can plausibly deny a client’s participation under a calibrated privacy parameter while maintaining competitive utility, yielding a clear outcome-level privacy interpretation. However, this requires careful calibration of the training and unlearning procedures.

Cross-user and ontology-guided defenses address leakage that arises from correlated features or representation neighbors. \textit{CUFRU}~\cite{li2025cross-user} which introduces a gradient transfer station and an iteration-aware calibration that blend historical and new gradients with time decay, enabling removal of a departing user while limiting spillover to neighbors in representation space. This improves privacy for cohorts but it requires careful control of the stored gradient signals. \textit{EG FedUnlearn together with OFU Ontology}~\cite{ghannam2025ontology-guided} adds structured knowledge to remove features linked to an entity consistently across holders and modalities while preserving benign attribute, reducing re-identification risk through correlated paths. This approach benefits from reliable ontologies and disciplined handling of small auxiliary metadata.

%Differential privacy and secure aggregation bound exposure at the mechanism and protocol levels. \textit{FedRecovery}~\cite{zhang2023fedrecovery} combines server-held historical information with differential privacy to produce unlearned models that approximate clean retains under a formal privacy budget, bounding leakage from both gradients and outputs at the cost of noise-utility trade-offs. \textit{Dynamic Participation with SecAgg+}~\cite{liu2024guaranteeing} preserves confidentiality when clients join and leave the federation by maintaining secure aggregation across changing sets so that no single update is revealed to the server during the erase or recovery phase. This strengthens protocol-level privacy at the edge, though it requires robust key management on constrained devices.
%
%Incentive- and policy-driven approaches add strategic reasoning to revocation. \textit{Strategic Data Revocation}~\cite{ding2024strategic} formalizes users’ revocation decisions as a non-cooperative game that balances privacy goals and utility, providing a principled way to characterize when and how users choose to revoke under regulatory and or incentive structures. Its effectiveness relies on accurate accounting of privacy impact and stable optimization under chosen policies.

Across these methods, the shared outcome is to make deletion results indistinguishable from a world where the target’s participation cannot be reliably inferred, while preventing adversaries from extracting sensitive facts from intermediate signals or final models. Attribute erasure and confusion-based forgetting limit inference from embeddings and activations, cross-user and ontology-guided designs suppress indirect leakage through correlated features, differentially private unlearning bounds exposure at the mechanism level, secure aggregation under dynamic participation hides individual updates during the procedure and game-theoretic revocation clarifies user incentives. Together, these choices align FUL at the mobile edge with strong privacy guarantees and practical utility.

\subsection{Cryptographic and blockchain protocols}
\label{subsec:crypto-blockchain}

Cryptographic and blockchain protocols make FUL \emph{verifiable} and \emph{tamper resistant} at the mobile edge. They attach audit trails to erase requests and outcomes, protect intermediate signals with encryption and anchor compliance claims to ledgers that independent parties can check. When designed carefully, these mechanisms reveal nothing about raw user data while keeping the unlearning workflow accountable under churn and adversarial behavior.

%blockchain-based accountability
\textit{BlockFUL}~\cite{liu2025blockful} redesigns blockchain-integrated FL so that unlearning actions are recorded and provable without requiring a full chain rewrite. It introduces a dual-chain architecture with a redactable live chain and an archive chain that preserves full history, supporting both parallel and sequential unlearning across inherited models. This ensures that certificates remain consistent along model lineages while guaranteeing immutability of audit trails and verifiable linkage between erase requests and updated model states and controlled redactions via redactable primitives such as chameleon hashes. Practical caveats include the need to protect trapdoor keys for redactable hashing, bound what metadata are written on the chains and manage committee-based authorization to avoid revealing deleted parties. \textit{BC-MU}~\cite{zuo2025trustworthy} complements this design by building a trustworthy framework that maintains a ledger of certified unlearning events tied to model versions, allowing operators to publicly prove that deletion requests were honored. While immutability and transparency enhance trust, this scheme leaves protection of raw client updates to cryptographic mechanisms off-chain and its efficiency depends on consensus settings adapted to edge deployments.

%encryption-driven revocation
\textit{RevFRF}~\cite{liu2022revfrf} develops an end-to-end encrypted protocol for federated random forests with participant revocation. Using homomorphic encryption, it supports secure construction, prediction and revocation so that the contribution of a revoked participant is securely removed without exposing other parties' samples. This provides clear revocation semantics for tabular workloads, though the limitations include the heavy computational overhead of homomorphic operations on constrained edge devices and the model specificity since the protocol targets tree ensembles rather than deep neural networks. 

%decentralized coordincation
Decentralized Unlearning for trustworthy Artificial Intelligence Generated Content (AIGC) services~\cite{lin2024decentralized} introduces a ledger-guided and coded computing architecture that orchestrates unlearning across multiple holders. By combining decentralized control with differential privacy, it ensures that no single party can reconstruct a user’s contribution while improving traceability and aligning with the RTBF in multi-owner generative pipelines. These benefits come with trade-offs in system complexity, careful policy design for coded interactions and the need to carefully calibrate privacy budgets against quality in generative AIGC workloads.

Together, these approaches demonstrate how cryptographic and blockchain tools extend FUL beyond local model edits to \emph{verifiable, auditable and accountable} workflows. Blockchains establish transparent and tamper-resistant records of unlearning events, homomorphic encryption enables mathematically secure revocation and decentralized protocols enforce trust across multi-party pipelines. The challenge lies in balancing strong verifiability with efficient performance and minimal metadata leakage at the resource-constrained mobile edge.

\subsection{Summary and Lessons Learned}
\label{sec:summary-security-privacy}

This section consolidates the insights gained from the surveyed works on security and privacy mechanisms for FUL at the mobile edge. Table~\ref{tab:security-privacy} summarizes representative approaches, and the key lessons, grounded in evidence from these studies, are discussed below.

\begin{itemize}
  \item \textbf{Poisoning and backdoor defense} methods such as \emph{FedBT}~\cite{Wang2025FedBT}, \emph{UnlearnGuard}~\cite{wang2025poisoning}, and \emph{robustFU}~\cite{sheng2024robust} demonstrate that integrating detection and filtering into the unlearning workflow substantially improves model integrity while reducing the recovery cost. These studies show that detection-guided aggregation and anomaly-aware distillation limit the propagation of poisoned gradients and recover clean performance in fewer rounds compared with baseline retraining. \emph{FedRemover}~\cite{yuan2024toward} and \emph{Get Rid of Your Trail}~\cite{alam2024trail} confirm that rollback and subtraction can remove malicious clients within a few local updates while preserving privacy through controlled metadata retention. \emph{FAST}~\cite{guo2023fast} and \emph{Backdoor KD}~\cite{wu2022knowledge} further support that server-side sanitization and distillation using proxy data can restore clean behavior without recalling all devices. Collectively, these results reveal that poisoning-resistant unlearning is feasible when reliable detection signals and compact rollback mechanisms are combined, although maintaining accuracy under non-IID data and avoiding false positives remain practical challenges.

  \item \textbf{Certified and verifiable unlearning} approaches including \emph{Starfish}~\cite{liu2025certifiedclientremoval}, \emph{CFRU}~\cite{huynh2025certified}, and \emph{CerFeaUn}~\cite{wang2025-feaun} provide concrete evidence that deletions have been correctly applied by producing certificates of removal based on statistical indistinguishability and reconstruction bounds. These works consistently demonstrate that verifiable proofs can replace subjective trust between participants and servers. For example, \emph{Starfish} achieves certified client removal through secure two-party computation, and \emph{CFRU} validates removal correctness in federated recommendation systems using sampled rollback verification. Similarly, \emph{VeriFi}~\cite{gao2024verifi} establishes end-to-end auditability by enabling users to test post-unlearning model responses without revealing private updates. Together, these systems confirm that verifiable deletion can be achieved with bounded metadata and formal proofs, though at the cost of additional computation and coordination to maintain lightweight verification in edge environments.

  \item \textbf{Privacy preservation and inference resistance} methods such as \emph{Aegis}~\cite{wu2025aegis}, \emph{ConFUSE}~\cite{meerza2024confuse}, and \emph{Plausible Deniability FU}~\cite{varshney2025plausible} show that carefully designed erasure of attributes or instances can substantially reduce membership and attribute inference risks without requiring access to full training data. \emph{CUFRU}~\cite{li2025cross-user} and \emph{EG FedUnlearn with OFU Ontology}~\cite{ghannam2025ontology-guided} demonstrate that cross-user calibration and ontology-guided removal suppress leakage across correlated representations and modalities, improving privacy consistency across holders. \emph{FedRecovery}~\cite{zhang2023fedrecovery} confirms that integrating differential privacy into unlearning keeps gradient and output exposure within defined bounds, while \emph{Dynamic Participation with SecAgg+}~\cite{liu2024guaranteeing} ensures confidentiality when clients join or leave during deletion through secure aggregation over dynamic membership sets. The overall lesson from these works is that privacy preservation requires a layered approach that combines mechanism-level noise, representation-level anonymization, and protocol-level concealment. The cited studies collectively reveal that such combinations can deliver measurable privacy gains with acceptable accuracy loss when noise and aggregation parameters are well-calibrated.

  \item \textbf{Cryptographic and blockchain protocols} such as \emph{BlockFUL}~\cite{liu2025blockful}, \emph{BC-MU}~\cite{zuo2025trustworthy}, and \emph{RevFRF}~\cite{liu2022revfrf} demonstrate that verifiable and tamper-resistant unlearning can be achieved when audit trails and encrypted operations are integrated into the workflow. \emph{BlockFUL} introduces a dual-chain ledger that records unlearning requests and results, ensuring traceability without exposing deleted data. \emph{BC-MU} reinforces this approach through certified ledgers that tie deletions to model versions, enhancing public accountability. \emph{RevFRF} extends confidentiality to encrypted tree models by allowing secure revocation under homomorphic encryption, showing that exact removal is possible even when data remain encrypted. Finally, decentralized designs such as \emph{Decentralized Unlearning for AIGC}~\cite{lin2024decentralized} combine privacy and traceability through coded computing and differential privacy, providing trustworthy orchestration across multiple data holders. Collectively, these works show that cryptographic and blockchain mechanisms can strengthen auditability and trust at the mobile edge, though their computational overhead and metadata management requirements remain the main obstacles for lightweight deployment.
\end{itemize}

Across these four families, a clear trend emerges. Systems that integrate security and privacy features early in the unlearning pipeline consistently achieve stronger guarantees with fewer additional rounds. Detection-guided workflows improve robustness before deletion, verifiable protocols provide measurable accountability after deletion, and privacy-preserving or cryptographic layers ensure confidentiality throughout the process. The combined evidence across all studies indicates that robust FUL at the mobile edge requires joint optimization of integrity, auditability, and confidentiality, with the most resilient systems blending multiple defenses rather than relying on any single family in isolation.

\section{Integrated Edge Constraints}
\label{sec:integrated-edge-constraints}

\begin{figure}
    \centering
    \includegraphics[width=0.8\linewidth]{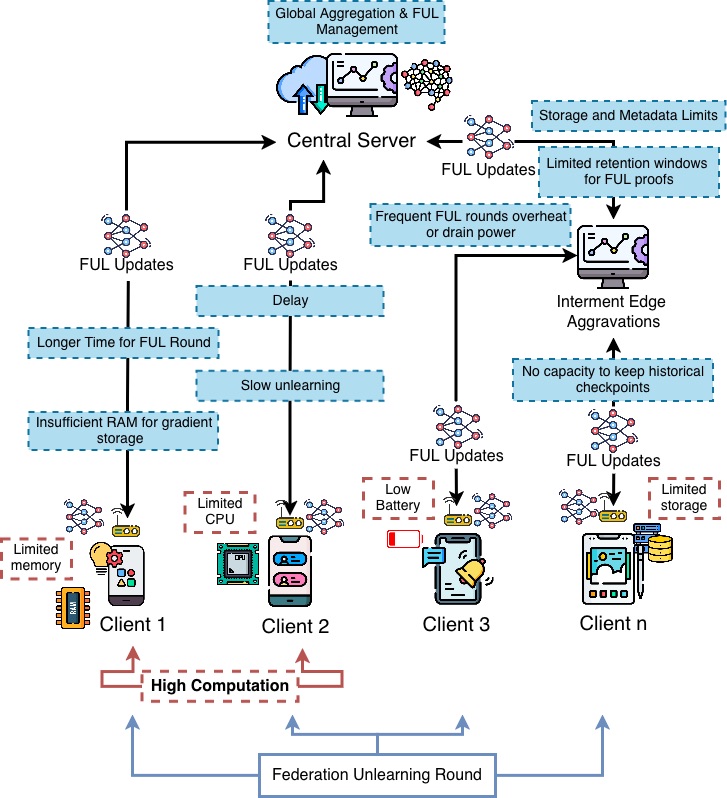}
    \caption{Resource Challenges of Federated Unlearning in Mobile Edge.}
    \label{fig:challangesresource}
\end{figure}

\begin{table*}[!t]
\centering
\caption{\small Classification of Federated Unlearning methods by Integrated Edge Constraints with Strategies, References, Core Mechanisms, Advantages, and Limitations.}
\label{tab:edgeconstraints}
\setlength{\tabcolsep}{4pt}
\begin{tabular}{|>{\centering\arraybackslash}m{1.3cm}|>{\centering\arraybackslash}m{1.3cm}|>{\centering\arraybackslash}m{1.6cm}|m{4.3cm}|m{3.8cm}|m{3.8cm}|}
\hline
\rowcolor{mygray}
\textbf{Subcategory} & \textbf{Strategy} & \textbf{References} & \textbf{Core Mechanism} & \textbf{Advantages} & \textbf{Limitations} \\
\hline

\textbf{Resource-aware and lightweight edge methods} &
Adapter based &
\textit{FedWiper}~\cite{zhao2025fedwiper}, \textit{FedEditor}~\cite{yuan2025fededitor} &
Small reversible adapters or editing modules modify a narrow subset of parameters so deletions are applied through compact deltas while the shared backbone remains stable. &
Communication stays low because only adapters are exchanged and edge devices finish deletion quickly with minimal energy. &
Adapter coverage may miss deeply entangled features and alignment across heterogeneous backbones can be challenging on diverse devices. \\
\cline{2-6}

& RL scheduling &
\textit{RL-FU}~\cite{gao2025reinforcement} &
A reinforcement learning controller selects the timing scope and device set for unlearning using latency energy and drift signals from the edge. &
Devices avoid redundant rounds and the system adapts to variable connectivity and power conditions. &
Policies require training data and may overfit to specific workloads and the controller adds runtime overhead on constrained nodes. \\
\cline{2-6}

& DP budgeted retention &
\textit{FedADP}~\cite{jiang2024daptivedifferential} &
Differential privacy budgets adapt during training so only impactful gradients are stored and redundant updates are discarded to keep minimal server state for deletion. &
Server storage remains small and privacy is preserved while retaining the ability to perform effective erasure. &
Noise budgets can reduce accuracy and budget misallocation can weaken deletion fidelity on rare classes. \\
\cline{2-6}

& On device forgetting &
\textit{FedUNRAN}~\cite{mora2024fedunran}, \textit{FedRollback}~\cite{ameen2024lightweight} &
Devices perform brief local forgetting such as random label steering or rollback then reconcile with a small server update. &
User level deletions proceed without federation wide retraining and battery usage remains acceptable. &
Local steps can create drift that requires careful reconciliation and very weak devices may still struggle with on device steps. \\
\cline{2-6}

& Heterogeneity tolerant KD &
\textit{HDUS}~\cite{ye2024heterogeneous} &
Distilled seed models align diverse local backbones so unlearning exchanges remain small and consistent under heterogeneous clients. &
Edge updates become compact and connectivity interruptions have less impact on progress. &
Distillation quality governs utility and seeds may not capture rare modalities which can reduce accuracy after deletion. \\
\hline

\textbf{Data quality and reliability management} &
Server filtering &
\textit{DIsFU}~\cite{kong2024disfu}, \textit{MetaFUL}~\cite{wang2023human-centric} &
Loss or bias based quality scores filter unreliable client updates and server side pseudo data impressions support deletion without full retraining. &
Large retraining cycles are avoided and corrupted data has limited influence on the global model. &
Quality scoring can remove useful signals and pseudo data can introduce distribution shift if it is not calibrated. \\
\cline{2-6}

& Credibility detection &
\textit{MCC-Fed}~\cite{wang2025malicious}, \textit{SIFU}~\cite{wang2024server-initiated} &
Credibility assessment identifies adversarial or low quality clients then triggers focused deletions or server initiated recovery phases. &
Benign contributions are preserved and communication overhead decreases because only targeted clients are contacted. &
Detectors can be evaded by adaptive behavior and false positives can penalize honest clients. \\
\cline{2-6}

& Local repair and fairness &
\textit{LMR}~\cite{ameen2024addressing}, \textit{F2UL}~\cite{su2024f2ul} &
Clients repair unreliable local models before aggregation and fairness aware optimization balances performance across groups during unlearning. &
Global utility improves and unaffected users maintain stable accuracy after deletion. &
Repair adds extra local computation and fairness objectives can slow convergence on scarce edge hardware. \\
\hline

\textbf{Heterogeneity and multi-modal adaptation} &
Vertical feature space &
\textit{VFL-Unlearning}~\cite{wang2024vertical}, V\textit{FU-KD}~\cite{varshney2025vertical}, \textit{FMU}~\cite{pan2025feature-based} &
Parties holding disjoint features coordinate through alignment layers or cross party distillation so user or attribute data is erased along relevant paths. &
Organizations remove sensitive features without sharing raw data and communication remains moderate through aligned interfaces. &
Cross party coordination increases complexity and misalignment can leave residual influence on composite features. \\
\cline{2-6}

& Graph and ranking data &
\textit{PAGE}~\cite{ai2025graph}, \textit{ReGEnUnlearn}~\cite{liu2025subgraph}, \textit{FU-FOLTR}~\cite{wang2024rank} &
Structure aware methods isolate and regenerate affected nodes subgraphs or ranking components to stabilize models after deletion. &
Topology is preserved and accuracy remains high in multi modal edge environments. &
Graph edits can be costly on large deployments and regeneration quality determines final utility. \\
\cline{2-6}

& Conflict aware optimization &
\textit{FedCSA}~\cite{yang2025fedcsa}, \textit{FedOSD}~\cite{pan2025conflict}, \textit{Ferrari}~\cite{gu2024ferrari}, \textit{FedLU}~\cite{zhu2023kgembedding} &
Clustering orthogonal projection and sensitivity reduction limit client interference so unlearning remains stable under non IID data. &
Training remains steady across diverse modalities and fewer rounds are needed to recover accuracy. &
Tuning projections and clusters requires care and sensitivity reduction can oversmooth features. \\
\hline

\textbf{Edge architecture and scheduling} &
Cache centric orchestration &
\textit{Edge Cache FU}~\cite{wang2024v2x}, \textit{FedSCC}~\cite{cui2024cluster-centri} &
Unlearning is localized to small cells or caches so only affected edge nodes update and backhaul traffic remains low. &
Service continuity improves at the edge and network congestion decreases during deletion. &
Cache consistency must be maintained and mislocalized updates can delay convergence. \\
\cline{2-6}

& Hierarchical async &
\textit{Hier-FUN}~\cite{ma2024hier-fun} &
Tiered coordination performs deletions within local clusters then reconciles asynchronously at higher tiers to reduce idle waiting. &
Latency decreases and other clusters continue training while a local deletion proceeds. &
Asynchronous reconciliation complicates bookkeeping and stale updates can reduce accuracy. \\
\cline{2-6}

& Model exchange &
\textit{FedME}~\cite{xia2023fedme} &
Compact model exchange replaces heavy gradient sharing so unlearning updates propagate quickly through the network. &
Network congestion is reduced and propagation time improves on bandwidth limited links. &
Model packaging requires careful versioning and exchange frequency must be tuned for stability. \\
\cline{2-6}

& Incentive aware scheduling &
\textit{Incentivized FL}\&U~\cite{ding2025incentivized} &
Pricing and reputation mechanisms encourage cooperative energy efficient participation and prioritize impactful deletions. &
Participation rates improve and costly round failures become less frequent in practice. &
Mechanism design increases protocol complexity and incentives may not generalize across deployments. \\
\hline
\end{tabular}
\end{table*}

FUL on the mobile edge faces fundamental resource challenges that go beyond algorithmic efficiency. Edge devices and small cells operate under tight budgets of compute, memory, and energy, while also dealing with variable data quality, intermittent connectivity, and evolving network topologies~\cite{lim2020fmobileedge, woisetschlager2024fledge, yan2024survey}. These constraints make it difficult to execute unlearning operations such as rollback, retraining, or correction without overloading local hardware or causing excessive delays in synchronization.

The problem comes because FUL introduces additional computation and communication steps compared to standard FL. Devices must store intermediate states, compute corrective updates, and occasionally rejoin after deletion, all of which strain already limited edge resources. Heterogeneous capabilities across devices further amplify this imbalance: powerful nodes can finish updates quickly, while weaker ones stall, leading to energy drain, idle waiting, and degraded model consistency.

Addressing these integrated edge constraints is therefore essential to make FUL both practical and sustainable. Figure~\ref{fig:challangesresource} summarizes the major bottlenecks caused by resource limitations and their cascading effects across devices and networks. To mitigate these issues, recent studies propose edge-aware designs that jointly consider computation, communication, and scheduling. These approaches aim to keep unlearning feasible on small devices, tolerate imperfect data and connectivity, and maintain accuracy while staying within strict energy and bandwidth budgets. The following families translate these ideas into concrete end-to-end designs that balance efficiency, scalability, and reliability in mobile-edge environments.

\begin{itemize}
  \item \emph{Resource-aware and lightweight edge methods:}
  These techniques shrink local computation, memory footprints and energy use so that unlearning remains viable on constrained edge devices. Techniques include pruning and quantitation to reduce model size, early exit and conditional computation to avoid unnecessary layers, split-style training~\cite{thapa2022splitfed} with compact feature uplinks and adapter or low-rank subnets to ensure that only a small parameter subset is touched during deletion. Local-step and message-size budgets are enforced to ensure opportunistic and battery-friendly participation~\cite{arouj2022battery}, which reduce on-wire traffic and shorten device wake times.
  
  \item \emph{Data quality and reliability management:}
  This family stabilizes unlearning under imperfect or drifting data by scoring and filtering updates before aggregation~\cite{pillutla2022robust} and maintaining lightweight reliability metadata instead of raw histories. It applies noise-aware training, confidence or influence weighting and outlier suppression to prevent rollback or targeted corrections from amplifying bad signals. Drift and fault detection trigger narrow recalibration rather than broad restarts while missing or stale contributions are handled with imputation or skip rules~\cite{karimireddy2020scaffold}, reducing re-transmissions and shortening recovery phase.

  \item \emph{Heterogeneity and vertical or multi modal adaptation:}
  These methods localize edits to personalized, feature-specific or modality-specific components so that only necessary slices of the model are updated during unlearning. Personalized heads and adapters allow clients to keep device- or domain-specific behavior while the shared core remains stable~\cite{karimireddy2020scaffold}, minimizing cross-client coordination. In vertical or feature-split settings, alignment layers~\cite{liu2024vertical} and secure aggregation ensure user removal edits only affects the relevant feature paths, while multimodal designs use lightweight modality adapters so that forgetting in one stream does not force full model update. These patterns identify who must participate and how much they must send.

  \item \emph{Edge architecture and scheduling:}
  Architecture and scheduler choices bound who talks and when under realistic connectivity and energy constraints. Hierarchical end-to-end layouts~\cite{abad2020hierarchical} confine most corrections to a nearby tier with infrequent reconciliation, asynchronous protocols~\cite{nguyen2022asynchronous} tolerate stragglers without stalling faster devices and bandwidth- or energy-aware schedulers prioritize clients whose contributions most affect the targeted deletion. Event-driven and opportunistic updates replace rigid rounds reducing idle synchronization and curbs repeated retries on flaky links.
\end{itemize}

\subsection{Resource-aware and lightweight edge methods}\label{sec:resource-aware-lightweight}

Resource-aware and lightweight edge methods keep unlearning feasible on small devices by shrinking compute and memory footprints and bounding how much each client must send or receive. They rely on compact edits such as adapters or subspaces, client-side forgetting that avoids federation-wide recall, dynamic participation that respects battery and link budgets and privacy controls that remove the need to retain heavy histories. The shared goal is to erase the target influence with short local routines and small messages so that compliance at the edge remains practical.

\textit{FedUNRAN}~\cite{mora2024fedunran} enables client-side unlearning by replacing true labels with random ones to steer the local model away from target signals, followed by a brief reconciliation with the server. This design avoids waking non-target devices and minimizes uplink usage while the client runs a short and simple routine that fits constrained hardware. Although this speeds up the unlearning process, there is a loss of accuracy on nearby classes, which requires careful tuning of randomization strength and recovery length so that utility remains stable under non-IID data distribution~\cite{mora2024fedunran}. \textit{FedRollback}~\cite{ameen2024lightweight} targets IoT sensing systems with a lightweight recovery pipeline that adapts to resource limits and uses simple aggregation for battery-friendly participation. While it limits wall-clock time, it introduces modest server-side bookkeeping for scheduling and monitoring~\cite{ameen2024lightweight}.

\textit{FedWiper}~\cite{zhao2025fedwiper} introduces a universal adapter that absorbs and then reverses the contribution to be forgotten while leaving the base network largely untouched. This converts deletion into a compact edit and avoids large model transfers~\cite{zhao2025fedwiper}. Its small adapter footprint and straightforward rollbacks suit edge devices tight storage and bandwidth, though maintaining adapter parameters and deciding when to refresh them to avoid drift on long training runs remain practical challenges. \textit{HDUS}~\cite{ye2024heterogeneous} addresses device heterogeneity by distilling seed models that serve as erasable building blocks, enabling alignment without heavy state sharing. This reduces synchronization payloads and tolerates diverse local backbones but performance depends on the quality of distilled seeds and stable teacher selection which can be challenging under intermittent participation~\cite{ye2024heterogeneous}.

%\textit{FedADP}~\cite{jiang2024daptivedifferential} saves resources by avoiding raw update logs through adaptive differential privacy with budget allocation that focuses only on impactful updates and selecting which stored items are worth keeping. This reduces storage and communication while protecting privacy, but balancing privacy noise against utility requires task-specific tuning~\cite{jiang2024daptivedifferential}. \textit{RL-FU}~\cite{gao2025reinforcement} applies deep reinforcement learning to decide when and how to unlearn under poisoning and drift while avoiding unnecessary computations and preserving model quality. This moves the decision burden from hand-tuned rules to a lightweight controller that saves energy and communication rounds at the edge.The policy training overhead and the challenge of stable exploration under non-stationary data distributions are its main limitations. \textit{FedEditor}~\cite{yuan2025fededitor} develops an efficient unlearning pipeline for cooperative intelligent transportation systems where mobile devices and fragile links demand that edits and exchanges focus on the smallest viable set. This minimizes compliance delays in practice, though savings rely on exploiting domain-specific structure.

Across these resource-aware designs, the common pattern is to keep edits local and compact while budgeting participation to only relevant clients, thereby lowering communication volume and shortening recovery. Client-side routines like random-label erasure remove target influence without federation-wide recall, adapters and distillation layers confine edits to small modules and privacy-guided selection avoids retaining or shipping bulky histories. Together, these methods deliver predictable latency and energy footprints on edge devices, making FUL practical for mobile-edge deployments.

\subsection{Data quality and reliability management}
\label{sec:data-quality-reliability}

Data quality and reliability management methods stabilize FUL under noisy labels, poisoned updates, stale models and intermittent clients. They operate by estimating the credibility of each client or update, filtering or down-weighting low-quality signals before aggregation, repairing or forgetting unreliable contributions with compact edits and triggering server-guided cleanups only when evidence crosses a threshold. The shared goal is to protect regular users and avoid broad retraining by keeping corrections narrow and reusing lightweight reliability metadata rather than storing heavy histories.

\textit{DIsFU}~\cite{kong2024disfu} reduces the involvement of innocent clients by performing unlearning primarily at the server with data impressions and bias-model correction, then restoring utility via knowledge distillation on pseudo-data. This method only requires the server and target client to participation. This limits privacy exposure and keeps recovery short, though its effectiveness depends on the fidelity of the pseudo-data and the stability of the bias correction. \textit{MCC-Fed}~\cite{wang2025malicious} jointly considers malicious-client detection and contribution quality, learning to distinguish harmful behavior from benign heterogeneity so that the server can scale or gate contributions accordingly. Its strength lies in focusing unlearning where it matters, though its success depends on reliably attributing harmful influence to avoid punishing benign but rare client patterns.

%\textit{LMR}~\cite{ameen2024addressing} identifies and categorizes unreliable local models and performs local refinement before aggregation so that the server receives cleaner signals for forgetting and recovery. This reduces recomputation and keeps payloads small, but requires careful calibration of thresholds to avoid excessive or unnecessary repair. \textit{F2UL}~\cite{su2024f2ul} formulates unlearning in data trading as a fairness-aware optimization problem that selects and calibrates removals to balance global utility and participant fairness. This improves the protection for regular users but introduces a tunable fairness-accuracy trade-off that must be tailored to the application context.

%\textit{MetaFUL}~\cite{wang2023human-centric} removes the impact of low-quality models at the server with a non-communicative unlearning routine informed by loss-based quality assessment. It improves robustness without additional client transmissions, but depends heavily on reliable quality estimators and careful server-side bookkeeping. \textit{SIFU}~\cite{wang2024server-initiated}, in the server-initiated sense, equips the server with a right-to-remove mechanism for low-quality local models and executes a narrow unlearning step followed by a brief recovery with only the necessary clients. This avoids round-heavy restarts but requires robust and auditable triggers for initiation to ensure accountability.

Across these approaches, the common pattern is to score, filter or repair updates before aggregating and to trigger narrow cleanups only when justified. Credibility weighting and server-initiated controls prevent noisy clients from dominating, while local repair and targeted forgetting shorten recovery and reduce synchronization overhead. The result is fewer retransmissions, smaller payloads and more stable accuracy for regular users under mobile-edge constraints.

\subsection{Heterogeneity and vertical or multi modal adaptation}
\label{sec:heterogeneity-multimodal}

Heterogeneity and vertical or multi-modal adaptation methods preserve unlearning accuracy when clients differ in data distributions, model capacity and feature spaces. They localize edits to personalized components, align features across parties that hold disjoint attributes and coordinate updates across modalities or structured objects so that only the slices of the system that carry the target signal are modified while the shared core remains stable. These strategies narrow who must participate and how much they must transmit, fitting the realities of bandwidth-limited and intermittently connected mobile-edge environments.

Vertical feature-space methods confine unlearning to the parties that store the relevant attributes, allowing other participants avoid large exchanges. \textit{VFL Unlearning}~\cite{wang2024vertical} removes a user or attribute's contribution by reversing the learning trajectory with a constrained gradient ascent procedure and validating the effect of the removal via a backdoor certification test, followed by a brief reconciliation narrower than federation-wide retraining. \textit{VFU KD}~\cite{varshney2025vertical} distills retained information across vertical partners so that the target signal is suppressed while communication remains light, since soft targets replace heavy gradients and only the active party’s stored embeddings are required. \textit{FMU}~\cite{pan2025feature-based} adopts a feature-based formulation of VFL for IoT networks so that edits target attributes rather than entire instances, reducing coordination overheads between different parties. \textit{FedLU}~\cite{zhu2023kgembedding} operates on federated knowledge graph embeddings where mutual knowledge distillation aligns heterogeneous clients during learning and the unlearning phase removes specific triples and propagates the effect to the global model while leaving unrelated embeddings stable.

Structured and multi-modal adaptations act on graphs, sequences and ranking pipelines where influence is concentrated in substructures. \textit{PAGE}~\cite{ai2025graph} provides a three-stage framework: (i) prototype matching for local removal, (ii) adversarial sample reconstruction, and (iii)negative knowledge distillation to purge cross-client permeation so that only affected partitions are updated. \textit{ReGEnUnlearn}~\cite{liu2025subgraph} addresses subgraph level requests by learning a reinforced sampling policy to isolate the neighborhood to edit and extracting client-specific knowledge for regeneration. This preserves utility while avoiding edits to the full backbone. \textit{FUL FOLTR}~\cite{wang2024rank} adapts online learning-to-rank systems by removing a user or item footprint in line with the online update rules while verifying the effect of deletion, maintaining ranking quality without broad retraining across modalities and features.

Conflict- and capability-aware controllers further stabilize accuracy under non-IID data and diverse device capacity. \textit{FedOSD}~\cite{pan2025conflict} computes orthogonal steepest-descent directions so that forgetting does not collide with retained optimization and the post-training recovery does not revert erased knowledge, shortening recovery phases under divergent data. \textit{FedCSA}~\cite{yang2025fedcsa} mitigates heterogeneity by clustering and slicing the user population and then aggregating selectively so that only silos with similar distributions collaborate, while unrelated silos stay untouched during a deletion, keeping traffic local to the affected group. Ferrari~\cite{gu2024ferrari} minimizes feature sensitivity using Lipschitz-based analysis so that edits are applied only to features that carry the target signal while leaving the shared backbone stable, thus reducing unnecessary synchronization across clients.

Across these designs, the unifying theme is adapting \emph{where} and \emph{how} unlearning occurs so that edits are applied in the right space and at the right granularity. Vertical approaches act on aligned features rather than entire models, structured and multi-modal designs act only on substructures carrying the signal, and conflict- or capability- aware controllers prevent unnecessary work on unrelated clients. The result is fewer participants per round, smaller payloads and short recovery phases with accuracy that closely match counterfactual retraining under heterogeneous mobile-edge conditions.

\subsection{Edge architecture and scheduling}
\label{sec:edge-arch-scheduling}

Edge architecture and scheduling methods restructure \emph{where} and \emph{when} unlearning takes place so that deletions remain localized and predictable under bandwidth and energy limits. They organize devices into tiers or clusters, place caches or coordinators close to users and run schedulers that select only the necessary participants at the right times. By confining edits to nearby nodes and reducing idle synchronization, these systems keep latency low and minimize bytes on the wire while preserving the fidelity of deletions.

\textit{Edge Cache FU}~\cite{wang2024v2x} and \textit{FedSCC}~\cite{cui2024cluster-centri} place cache-centric logic at small cells and use federated coordination to decide what to cache and what to forget. Upon a deletion request, the cell executes local unlearning and cache refresh so that only the affected cache and its clients exchange messages, leaving the backbone untouched. This design reduces round trips to the core network and contains errors introduced by stale or invalid content, which is particularly valuable for mobile users moving across cells. The approach works best when clusters are stable and cache hit patterns are predictable, with modest overhead for maintaining cache state and history at the edge.

\textit{Hier-FUN}~\cite{ma2024hier-fun} organizes devices into hierarchical tiers where cluster heads aggregate and unlearn within their groups before sending compact updates upward. This confines most traffic to the lower tier and allows upper layers to reconcile occasionally, shortening recovery time under intermittent links. The benefits rely on selecting an appropriate number of clusters and capable heads, as well as balancing data heterogeneity across tiers so that local corrections do not drift. \textit{FedME}~\cite{xia2023fedme} complements this architecture by redefining and provisioning the exchange unit at the edge. In particular, instead of heavy gradients, the clients swap lightweight models or surrogates. Exchanging compact models and scheduling who swaps with whom reduce congestion across hotspots while requiring careful compatibility handling across clients with heterogeneous backbones.

\begin{figure}
    \centering
    \includegraphics[width=1\linewidth]{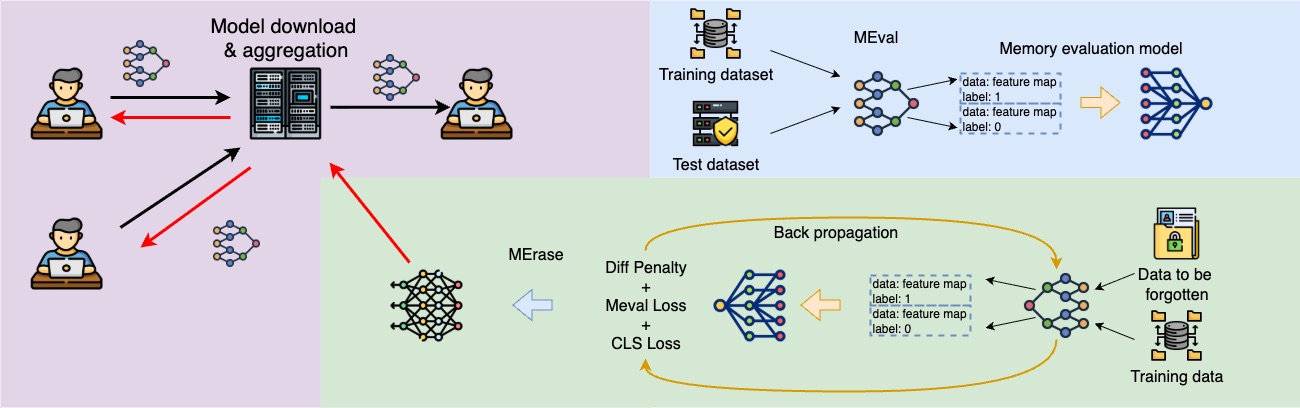}
    \caption{High Level Architecture of \textit{FedME}.}
    \label{fig:fedme}
\end{figure}

\textit{Incentivized FL and Unlearning} integrate mechanism design into scheduling so that valuable clients remain engaged and avoid unnecessary removals that trigger heavy cleanups~\cite{ding2025incentivized}. Pricing and reputation signals influence participation and unlearning decisions so that the system prioritizes clients that improve utility, while delaying or compensating removals that would otherwise impose large network costs. From a security and privacy perspective, such incentive schemes can discourage malicious or free-riding behavior, reduce the incentive to launch poisoning attacks that later require expensive unlearning, and provide an auditable record of who requested deletions and why. At the same time, the mechanism must protect the privacy of payment and reputation information, and its control plane must be robust against strategic manipulation and fair across diverse device classes.

Across these approaches, the central these is \emph{structural containment}: organizing who participates, when reconciliation occurs and how exchanges are represented so that unlearning traffic stays narrow and predictable. By pushing logic into caches, clusters or schedulers, systems localize deletions, cut backbone usage and stabilize compliance costs in mobile-edge environments.

\subsection{Summary and Lessons Learned}
\label{sec:summary-edge-constraints}

This section consolidates the insights drawn from the surveyed studies on integrated edge constraints for federated unlearning. Table~\ref{tab:edgeconstraints} summarizes representative methods, and the following observations are grounded in the evidence provided by these works.

\begin{itemize}
  \item \textbf{Resource-aware and lightweight methods} such as \emph{FedUNRAN}~\cite{mora2024fedunran}, \emph{FedRollback}~\cite{ameen2024lightweight}, and \emph{FedWiper}~\cite{zhao2025fedwiper} demonstrate that model pruning, adapter-based editing, and random-label erasure can keep unlearning feasible even on small and energy-limited devices. The results reported by these works show that local memory use and communication volume can be reduced substantially while maintaining accuracy that is close to retraining from scratch. Later studies including \emph{FedADP}~\cite{jiang2024daptivedifferential} and \emph{RL-FU}~\cite{gao2025reinforcement} confirm that adaptive privacy budgeting and reinforcement-based scheduling can further save energy and communication by activating clients only when necessary. Together, these works reveal that resource-aware unlearning is effective when model compactness, energy budgets, and training fidelity are jointly optimized, although some loss in precision may appear when the models become too small or the privacy budgets are overly restrictive.

  \item \textbf{Data quality and reliability management} approaches such as \emph{DIsFU}~\cite{kong2024disfu}, \emph{MCC-Fed}~\cite{wang2025malicious}, and \emph{MetaFUL}~\cite{wang2023human-centric} show that filtering and reweighting unreliable updates before aggregation can stabilize accuracy under noisy or drifting data. These works collectively indicate that reliability scoring and loss-based filtering help the system to identify harmful or low-quality updates, reducing the need for expensive retraining and shortening recovery phases. \emph{SIFU}~\cite{wang2024server-initiated} extends this idea by allowing the server to initiate focused cleanups only when certain quality conditions are met, preventing unnecessary synchronization. The overall lesson from these studies is that the stability of unlearning depends more on the quality and timing of selected updates than on their quantity, provided that the credibility estimators and thresholds are properly calibrated.

  \item \textbf{Heterogeneity and vertical or multimodal adaptation} methods including \emph{VFL Unlearning}~\cite{wang2024vertical}, \emph{VFU KD}~\cite{varshney2025vertical}, and \emph{PAGE}~\cite{ai2025graph} illustrate that aligning model components by feature, modality, or subgraph can confine deletions to a narrow portion of the model while preserving global utility. These approaches report consistent improvements in communication efficiency and recovery time, particularly in non-IID and multi-domain environments. Works such as \emph{FedCSA}~\cite{yang2025fedcsa} and \emph{FedOSD}~\cite{pan2025conflict} further demonstrate that conflict-aware and cluster-aware coordination reduces redundant participation and limits cross-client interference. The consistent evidence from these studies indicates that adapting unlearning to the heterogeneity of data and model structure is essential for scalable and accurate operation at the mobile edge, though maintaining stable alignment across modalities and silos remains an open challenge.

  \item \textbf{Edge architecture and scheduling} frameworks including \emph{Edge Cache FU}~\cite{wang2024v2x}, \emph{Hier FUN}~\cite{ma2024hier-fun}, and \emph{FedME}~\cite{xia2023fedme} demonstrate that reorganizing communication through caching, clustering, and hierarchical coordination significantly reduces latency and backbone traffic during deletion. Their findings show that when local caches or cluster heads handle unlearning within their tier, most corrections can be performed without involving the entire federation. The addition of incentive-aware mechanisms such as in \emph{Incentivized FL and Unlearning}~\cite{ding2025incentivized} highlights that participation can be guided toward high-utility actions by linking resource use to contribution value. Across these studies, the consistent observation is that architectural containment and adaptive scheduling make unlearning more predictable under bandwidth and energy constraints, although they introduce extra complexity in coordination, fairness management, and control logic.
\end{itemize}

When these families are viewed together, a clear pattern emerges. The most successful systems address multiple constraints at once, combining lightweight model design with reliability scoring or integrating hierarchical scheduling with adaptive resource management. Such hybrid strategies achieve stable trade-offs between accuracy, latency, and energy consumption across diverse mobile-edge environments. The overall lesson is that communication efficiency and computational feasibility in federated unlearning arise from the joint management of device capacity, data reliability, client diversity, and network structure, rather than from optimizing any single constraint in isolation.

\section{Applications of Federated Unlearning}
\label{sec:applications-ful-mec}

\begin{table*}[!ht]
\caption{\small Representative mobile-edge applications combining FUL with edge capabilities.}
\label{tab:apps-ful-edge}
\small
\centering
\setlength{\tabcolsep}{4pt}
\renewcommand{\arraystretch}{1.1}
\begin{tabular}{|>{\centering\arraybackslash}m{1.5cm}|>{\centering\arraybackslash}m{4.2cm}|m{10.5cm}|}
\hline
\rowcolor{mygray}
\textbf{Application} & \textbf{References} & \multicolumn{1}{c|}{\textbf{Description}} \\
\hline

Cyberattack Recovery &
\textit{FedRemover}~\cite{yuan2024toward}, \textit{FAST}~\cite{guo2023fast}, \textit{Backdoor KD}~\cite{wu2022knowledge}, \textit{UnlearnGuard}~\cite{wang2025poisoning}, \textit{Get rid of your trail}~\cite{alam2024trail} &
Rollback or subtraction removes malicious clients/windows; lightweight sanitization suppresses triggers; teacher--student transfer preserves benign utility; hardened deletion prevents erase-time exploits. \\
\hline

Consent-aware Cache \& Offloading &
\textit{Edge Cache FU}~\cite{wang2024v2x}, \textit{FedSCC}~\cite{cui2024cluster-centri}, \textit{FedME}~\cite{xia2023fedme}, \textit{FedAU}~\cite{gu2024unlearningduringlearning}, \textit{FednP}~\cite{jia2025fednp}, \textit{FedUMP}~\cite{zhu2024federated} &
Cluster-scoped deletion confines erasures spatially; compact model exchange reduces refresh cost; in-training unlearning enables prompt revocation; partition-level orchestration bounds reconciliation. \\
\hline

Personalized On-Device AI &
\textit{FedWiper}~\cite{zhao2025fedwiper}, \textit{FUSED}~\cite{zhong2025unlearning}, \textit{SFU}~\cite{li2023subspace}, \textit{QuickDrop}~\cite{dhasade2024quickdrop}, \textit{FedUHB}~\cite{jiang2024feduhb}, \textit{FedU}~\cite{wang2024fedu} &
Localized adapters and subspace edits enable instant retracts; synthetic exemplars support data-light fixes; heavy-ball and influence approximation reduce erase rounds on constrained devices. \\
\hline

Vehicular Networks &
\textit{DT-FU}~\cite{daluwatta2024dt-fu}, \textit{UaaS-SFL}~\cite{daluwatta2024uaas}, \textit{FedEditor}~\cite{yuan2025fededitor} &
Digital twins validate deletions pre-deployment; split unlearning shifts heavy steps to edge/server; segment-level edits ensure fast, verifiable RTBF compliance under V2X constraints. \\
\hline

\end{tabular}
\end{table*}

In the aforementioned studies, we have examined the challenges surrounding FUL as an enabling capability for at the mobile edge. In this section, we focus on the applications of FUL in mobile-edge network optimization and user-facing services. Prior work in mobile edge computing and FL has shown that the heterogeneity and complexity of wireless systems motivate data-driven control and policy learning at the edge. Yet, the underlying data is sensitive and client participation is intermittent. FUL extends FL by enabling systems to \emph{remove} specific users, items, features or time windows from trained models through narrow corrections and short recovery. This allows compliance with right-to-erasure obligations, restores model integrity and preserves performance under edge constraints. We highlight four representative application areas:

\subsection{Cyberattack Recovery}
\label{subsec:cyberattack-recovery}

Mobile edge environments such as vehicular networks, IoT infrastructures, and healthcare gateways are increasingly exposed to adversarial activity, including model poisoning and backdoor injections from compromised clients or edge applications. FUL provides an essential post-incident recovery mechanism that complements conventional intrusion detection and mitigation pipelines. Rather than retraining models from scratch after an attack, the system can selectively remove the influence of malicious updates, restoring a trustworthy global model while preserving data privacy and operational continuity.

In a typical recovery workflow, edge operators first deploy attack-aware detectors and robust aggregation mechanisms to identify suspicious contributions or poisoned patterns. Detection techniques such as spectral signature discovery and neural activation screening~\cite{wang2019neural, tran2018spectral} are combined with robust aggregation rules that bound the propagation of malicious gradients~\cite{fung2018mitigating, cao2020fltrust}. Once a set of adversarial clients or updates is confirmed, FUL is invoked to excise their residual influence and recalibrate the model with minimal disruption to normal edge operations. FUL based recovery operates through several complementary modes that correspond to different stages of the remediation pipeline. 
\begin{enumerate}[label=(\roman*)]
\item \emph{Targeted rollback}, which subtracts or cancels logged adversarial contributions before performing a lightweight clean pass to recover accuracy. 
\item \emph{Sanitization}, which weakens or overwrites poisoned features using distilled knowledge from a verified teacher model. 
\item \emph{Server-side correction}, which leverages stored statistics or checkpoints to complete deletion without recalling benign devices.
\end{enumerate}

These operational modes have been realized through concrete systems. \emph{FedRemover}~\cite{yuan2024toward} reconstructs a clean model by reversing the attacker’s gradient trace while maintaining an auditable trail for forensic review. \emph{FAST}~\cite{guo2023fast} performs rapid backdoor sanitization on the server to suppress trigger pathways, and \emph{Backdoor KD}~\cite{wu2022knowledge} transfers benign behavior from safe teacher models to remove malicious signatures without client re participation. To prevent attackers from exploiting the unlearning phase itself, \emph{UnlearnGuard}~\cite{wang2025poisoning} integrates screening and consistency checks into the FUL workflow, ensuring that only verified corrections are accepted.

Recent architectural extensions strengthen this recovery process in practical deployments. \emph{Unlearning as a Service}~\cite{daluwatta2024uaas} executes sensitive steps near the edge or server, reducing attack surfaces and enabling policy enforcement under regulated settings such as industrial IoT and healthcare. Digital-twin–driven orchestration~\cite{daluwatta2024dt-fu} validates deletions within a simulated twin environment before synchronizing updates to production, ensuring safe rollback without service interruption. Fairness-aware and attack-adaptive objectives~\cite{sheng2024robust} enhance resilience under imbalanced or adversarial cohorts, while trust-based aggregation such as \emph{FedBT}~\cite{Wang2025FedBT} helps isolate and down-weight corrupted clients before FUL is triggered. Finally, footprint-stripping approaches~\cite{alam2024trail} formalize the removal of attacker traces from the training record, achieving near-retrain performance while maintaining verifiable provenance for audit and compliance.

Across these designs, the integration of FUL within cyberattack recovery workflows transforms unlearning from a theoretical mechanism into an operational tool for real-time resilience. By coupling attack detection, trustworthy aggregation, and auditable deletion, mobile-edge systems can excise compromised contributions quickly and restore a clean, privacy-preserving model without downtime. This alignment between security response and FUL demonstrates FU’s direct practical relevance for maintaining reliability, accountability, and service availability in large-scale, heterogeneous edge environments.

\subsection{Consent-aware Cache and Offloading}
\label{subsec:consent-aware-cache}

Edge caching and offloading systems learn user preferences and content-placement policies from interaction traces collected near small cells and roadside units. These traces capture who requested what, when, and where, creating a clear privacy obligation when user consent changes or specific content is withdrawn. FUL extends these systems with the ability to revoke this information efficiently, ensuring that personalized caches and offloading models remain compliant without halting network operations.

In a typical deployment, each small cell or roadside unit participates in federated training to optimize cache placement and task-offloading decisions based on local traffic patterns. When a user withdraws consent or a content item is removed, the corresponding edge cluster must update its learned policies while avoiding costly federation-wide retraining. Recent designs integrate FUL directly into the cache control pipeline so that only affected clusters refresh their models, leaving the backbone undisturbed~\cite{wang2024v2x,cui2024cluster-centri}. In these systems, the controller identifies the specific user, item, or time window associated with the revocation and triggers a localized unlearning step that erases the relevant parameters, followed by a short recovery within that cluster. This workflow aligns the legal requirement for consent withdrawal with an operational mechanism for selective model editing.

To keep the refresh process efficient, communication and computation costs are minimized. Heavy gradient exchanges are replaced with compact model transfers~\cite{xia2023fedme}, and lightweight erase routines are embedded directly into the ongoing training loop so that new consent states propagate immediately~\cite{gu2024unlearningduringlearning}. The scheduling of edge participants is further optimized to maintain high cache hit rates and low latency while staying within backhaul and energy budgets. Asynchronous cluster orchestration allows unaffected clusters to continue normal learning~\cite{su2023asynchronous}, and partition-scope selection limits reconciliation to partitions that contain revoked data~\cite{jia2025fednp,zhu2024federated}. 

Through this integration, FUL becomes an operational component of consent management rather than a post-hoc correction. Cache and offloading controllers can now adapt in real time to user withdrawals or regulatory takedowns, ensuring that removed data no longer influences placement or routing decisions. The resulting system delivers localized, traffic-efficient, and latency-predictable updates that maintain both compliance and service continuity at the mobile edge.

\subsection{Personalized On-Device AI}
\label{subsec:personalized-ondevice}

Modern phones, wearables, and cameras increasingly rely on personalized AI models that learn from individual usage patterns to improve prediction, recommendation, and interaction. Examples such as voice and keyboard suggestions on mobile devices already demonstrate the feasibility of federated large-scale learning on a device, where models are updated locally to preserve privacy while improving utility on millions of devices~\cite{hard2019keyboard}. However, as personalization deepens, users also expect the ability to retract their data or reset individual models instantly. FUL extends on-device personalization into a revocation-aware framework, allowing rapid withdrawal or correction of user-specific knowledge without recalling or retraining every device.

In this application setting, FUL acts as the operational bridge between personalization and consent control. Adapter-based overwriting enables each device or the coordinating server to reverse user-specific characteristics using compact modules rather than full model updates~\cite{zhao2025fedwiper,zhong2025unlearning}. This design supports fast `undo` operations, ensuring that user traits, preferences, or biometric patterns can be erased on demand with minimal communication. Subspace editing also localizes forgetting to an orthogonal subspace derived from retained users so that only a small portion of parameters are adjusted during unlearning~\cite{li2023subspace}, stabilizing the shared model and preventing drift between heterogeneous devices. When direct access to raw data is unavailable, teacher–student transfer and synthetic exemplars~\cite{dhasade2024quickdrop} serve as safe surrogates for quick on-device corrections, maintaining personalization quality while minimizing computation and bandwidth during the retraction update cycles. Latency can be further reduced with explicit accelerators, such as heavy ball dynamics, that trace counterfactual solutions in fewer steps~\cite{jiang2024feduhb}, and with in-training unlearning that embeds lightweight erase paths directly into the local update loop so that retractions occur seamlessly as models learn~\cite{gu2024unlearningduringlearning}. Influence approximation strategies ensure that only the requesting device participates in deletion, minimizing collateral communication, and preserving scalability~\cite{wang2024fedu}.

From a systems perspective, these personalization workflows are reinforced by privacy-preserving protocols. Secure aggregation prevents the server from inspecting individual correction updates~\cite{bonawitz2017secure_agg}, and adaptive clipping with privacy accounting keeps information exposure bounded across continual revisions~\cite{xie2024adaptive}. The mature FL deployment patterns already used in production~\cite{bonawitz2019towards} simplify the transfer of small adapters or LoRA deltas across various hardware, allowing mobile and wearable devices to maintain synchronized but private models under fluctuating connectivity.

By combining parameter-efficient personalization with FUL primitives such as adapters, subspace edits, surrogate-based corrections, and accelerators, on-device AI systems can support immediate and privacy-preserving retractions alongside continuous personalization. This integration transforms FUL from a maintenance tool into an interactive feature, enabling users to update or withdraw their models on demand while keeping accuracy, latency, and privacy within the tight constraints of real-time mobile-edge environments~\cite{houlsby2019parameter,hu2022lora}.

\subsection{Vehicular Networks}
\label{subsec:vehicular-networks}

Vehicular edge systems continuously learn from real-time traffic flows, perception data, and charging-demand patterns collected through roadside units (RSUs) and connected vehicles. These applications operate under strict latency constraints and highly dynamic connectivity, which make conventional retraining infeasible when data must be revoked. Federated unlearning offers a practical revocation mechanism for this environment, enabling vehicles or fleet operators to erase trajectories, sensor intervals, or driver-related attributes without requiring raw data collection or federation wide retraining.

In a typical scenario, a driver invokes the RTBF or a fleet operator requests the removal of faulty or privacy-sensitive sensor segments. The vehicular edge controller identifies the corresponding feature traces and triggers FUL within the affected region. Instead of restarting the global model, the RSU performs a brief local recovery that corrects the aggregated state, allowing the rest of the network to continue normal operation. This allows personalized, regulation-compliant adaptation while maintaining low latency across V2X links.

To ensure reliability under mobility and churn, FUL in vehicular networks is orchestrated using hybrid control architectures. A digital-twin framework stages deletions in a virtual replica of the environment, validating the effect of erasure before synchronizing lightweight corrections with production systems~\cite{daluwatta2024dt-fu}. This enables verifiable and low-risk deletion even as vehicles and RSUs move in and out of coverage. In parallel, unlearning-as-a-service for split federated learning executes sensitive unlearning steps near the edge or central servers, keeping private computations in controlled environments while vehicles exchange only compact intermediate features~\cite{daluwatta2024uaas}. Both designs align FUL with the operational realities of vehicular infrastructure, where resource limits, connectivity gaps, and safety requirements demand predictable and isolated updates.

Application-specific pipelines further adapt FUL to transportation workloads. For example, edge editors such as \emph{FedEditor}~\cite{yuan2025fededitor} localize deletions to the smallest viable segment of the model so that RSUs update only those parameters influenced by the revoked data, restoring utility quickly without disturbing other regions of the network. Cache-centric and cluster-centric small-cell frameworks~\cite{wang2024v2x,cui2024cluster-centri} extend this idea to vehicular offloading and content delivery by refreshing only the caches or policies associated with the affected corridor rather than retraining the entire system. This locality-aware execution reduces backhaul load, limits the number of active clients per round, and sustains smooth service as vehicles move across cells.

Through this integration of digital-twin orchestration, split execution, and cluster-local cache refresh, FUL becomes an operational component of vehicular intelligence rather than an offline correction step. Edge systems can now revoke trajectories, attributes, or user identifiers rapidly and verifiably, maintaining compliance with privacy regulations while preserving service continuity. By aligning the mechanics of unlearning with the communication and safety constraints of V2X networks, FUL directly enhances both the resilience and trustworthiness of connected transportation infrastructures.

\section{Open Challenges and Future Directions}
\label{sec:open_challenges}

\paragraph{\textbf{Availability, asynchrony and dropouts}}
FUL must remain robust under intermittent connectivity and device churn without reopening large retraining windows. Buffered or fully asynchronous schemes sustain progress when clients arrive late or go offline, as demonstrated in buffered aggregation and asynchronous optimization for FL~\cite{nguyen2022asynchronous}. Practical schedulers should prioritize clients that most accelerate time-to-compliance under resource constraints, extending client selection ideas from mobile-edge FL to the unlearning context~\cite{nishio2019client}. From a broader systems view, deletion workflows benefit from staleness-aware rules and admission-control primitives tailored to asynchrony~\cite{kairouz2021advances}.

\paragraph{\textbf{Heterogeneity, personalization, and fairness}}
Non-IID data and uneven device capabilities can cause unlearning to disproportionately impact minority users or rare domains. Proximal and variance-reduction techniques, effective against client drift in standard FL, can be adapted to stabilize post-deletion recovery~\cite{karimireddy2020scaffold}. Beyond stability, fairness-aware formulations can increase the bound loss in disadvantaged clients by weighting the objectives during and after unlearning to mitigate accuracy gaps~\cite{li2019fair}. Survey evidence further highlights the need for routine disparity audits after each removal to ensure equitable outcomes in practice~\cite{kairouz2021advances}.

\paragraph{\textbf{Privacy beyond deletion}}
Even after removal, traces of membership or attributes may persist in outputs, activations, or gradients. Mechanisms such as Rényi DP provide refined composition for unlearning under privacy budgets~\cite{mironov2017renyi}, while classical DP training offers baselines for acceptable noise scheduling during erase and recovery~\cite{abadi2016deep}. Protocol-level protections shield intermediate updates under churn, making secure aggregation a natural foundation for privacy-preserving FU~\cite{bonawitz2017secure_agg}. Empirical membership-inference tests give concrete post-deletion checks for leakage in practice~\cite{carlini2022membership}.

\paragraph{\textbf{Verifiability, provenance and audit}}
Operators and regulators will demand evidence that a user, feature or slice has truly been removed. Proof-of-learning and proof-of-training approaches bind outcomes to training traces, suggesting how to construct auditable FUL logs without exposing raw data~\cite{jia2021proof}. Complementary documentation standards, including datasheets~\cite{gebru2021datasheets} and model cards~\cite{mitchell2019model}, provide the metadata necessary to resolve requests and report residual risks.

\paragraph{\textbf{Cryptography and trusted execution for erase pipelines}}
Certain deletions require cryptographic guarantees despite edge constraints. Secure and practical aggregation variants protect per-client updates under dynamic participation~\cite{bonawitz2017secure_agg}. Trusted Execution Environments (TEEs) isolate sensitive steps of the erase pipeline when bandwidth or energy preclude heavier cryptography~\cite{ohrimenko2016oblivious}. For verifiable retention or policy-driven redaction, lightweight tools such as approximate homomorphic encryption and redactable ledgers offer additional assurances~\cite{cheon2017homomorphic}.

\paragraph{\textbf{Communication-compute co-design}}
Uplink limits threaten to bottleneck FUL unless correction and recovery are explicitly co-optimized for compression. Quantization and sparsification reduce payload sizes with negligible accuracy loss when paired with error feedback, which is especially crucial during short recovery phases~\cite{lin2017deep}. Future unlearning primitives should expose low-rank or adapter-style interfaces so that small deltas can be applied and rolled back at predictable cost in edge environments.

\paragraph{\textbf{Concept drift, continual erasure and long-horizon stability}}
As edge distributions evolve with time and context, FUL must coexist with drift detection and adaptation. Classical drift monitors can trigger erase events or extra calibration only when needed~\cite{gama2014survey}, while reviews of drift adaptation recommend budgeted responses to avoid destabilizing retained knowledge~\cite{lu2018learning}. Federated surveys emphasize the importance of tracking long-term metrics such as recontamination risk and outcome stability across repeated deletions~\cite{kairouz2021advances}.

\paragraph{\textbf{Evaluation, datasets and reproducibility}}
FUL requires benchmarks that combine deletion streams, non-IID drift and intermittent participation. Existing FL suites such as LEAF~\cite{caldas2018leaf} and FedScale~\cite{lai2022fedscale} already provide mobile traces and large-scale tooling that can be extended with privacy probes and specific unlearning verifiability checks~\cite{caldas2018leaf,lai2022fedscale}. Recent evaluations of unlearning in large models underline the value of indistinguishability metrics and transfer robustness, which should be ported into mobile edge scenarios~\cite{geng2025comprehensive}.

\paragraph{\textbf{Scheduling, incentives and economics}}
Erasure incurs costs in bandwidth, energy and post-deletion utility. Guided participant selection can prioritize clients that accelerate cleaning time while respecting the limits of the device~\cite{lai2021oort}. Broader surveys on client selection suggest incentive-compatible policies that discourage strategic behavior yet respecting deletion rights in marketplaces and shared infrastructures~\cite{fu2023client}. Practical implementations should expose these signals to the unlearning scheduler.

\paragraph{\textbf{Governance and regulatory alignment}}
Mobile-edge FUL must comply with erasure rights and AI risk management guidelines. The GDPR’s right to erasure motivates machine-checkable policies and traceable logs that persist across model tiers~\cite{GDPR}. Risk frameworks such as NIST AI RMF~\cite{nist_ai_rmf_2023} and the EU AI Act~\cite{act2024eu} require controls and audit trails that persist through model updates and unlearning events.

\paragraph{\textbf{Systematization across the stack}}
End-to-end blueprints are needed to integrate asynchronous transport, privacy accounting, robust aggregation, cryptographic or TEE protections and verifiability into unified erase pipeline. At-scale FL system designs demonstrate workable orchestration and storage strategies that FUL can extend to heterogeneous edge deployments~\cite{bonawitz2019towards}. Community reference implementations will be essential for reproducibility, comparability and accelerated adoption in real deployments~\cite{kairouz2021advances,lai2022fedscale}.

\section{Conclusion}
\label{sec:conclusion}

This paper has presented a tutorial of federated unlearning (FU) and a comprehensive survey on issues regarding its implementation at the mobile edge. Firstly, we begin with an introduction to the motivation for mobile edge computing and why FUL must be a first-class capability alongside federated learning to meet legal and user-driven deletion needs. Then, we describe the fundamentals of unlearning goals, threat and trust models, and system design principles toward scalable, edge-ready FU. Afterwards, we provide detailed reviews, analyses, and comparisons of approaches across four core families, provable and server-side exactness, selective and rollback-based strategies, clustering/partitioning and participation control, and compression, distillation, and acceleration. Furthermore, we also discuss integrated edge constraints that cut across methods, including resource-aware and lightweight designs, data quality and reliability management, heterogeneity and multi-modal or vertical adaptation, and edge architecture and scheduling, as well as security and privacy topics such as poisoning \& backdoor defense, certified and verifiable unlearning, privacy preservation and inference resistance, and cryptographic or ledger-backed protocols. Finally, we discuss open challenges and future research directions spanning continual and asynchronous unlearning, fairness under non-IID data, verifiability and provenance, protocol hardening, incentive-aware scheduling, and reproducible evaluation at scale.

\bibliographystyle{ieeetr}
\bibliography{federatedunlearning}

\end{document}